\documentclass[final,3p,times]{elsarticle}

\journal{***}

\usepackage{float} 
\usepackage{amsmath}
\usepackage{mathtools}
\usepackage{amssymb}
\usepackage{subcaption}
\usepackage{tabularx}
\usepackage{color}
\usepackage[export]{adjustbox}
\renewcommand{\vec}[1]{\underline{\boldsymbol{#1}}}
\newcommand{\ten}[1]{\underline{\underline{\boldsymbol{#1}}}}

\usepackage{cases}
\everymath{\displaystyle}

\newcommand{\dovartilde}[2]{%
  \ifx#1\displaystyle\widetilde{#2}\else\tilde{#2}\fi
}

\usepackage{url,color,bm,lineno,xspace,soul,ulem}
\definecolor{mygreen}{rgb}{0,0.7,0.3}
\definecolor{myblue}{rgb}{0,0.3,0.8}
\definecolor{myred}{rgb}{0.65,0.2,0}
\definecolor{darkmagenta}{rgb}{0.65,0.2,0}

\definecolor{mygreen1}{rgb}{0.09,.45,0.27}

\usepackage{ifthen}
\newboolean{correction_mode}
\setboolean{correction_mode}{true}
\ifthenelse{\boolean{correction_mode}}
{
    \newcommand{\jvrm}[1]{{\color{myred}\sout{#1}}}
    \newcommand{\vyrm}[1]{{\color{myred}\sout{#1}}}

    \newcommand{\jvcm}[1]{{\color{mygreen}JV: #1}}
    \newcommand{\vycm}[1]{{\color{mygreen}VY: #1}}
    
    \newcommand{\abrm}[1]{{\color{red}\sout{#1}}}

\newcommand{\abrcm}[1]{{\color{mygreen1}\textbf{ABR: #1}}}

}
{
    \newcommand{\jvrm}[1]{}
    \newcommand{\vyrm}[1]{}

    \newcommand{\jvcm}[1]{}
    \newcommand{\vycm}[1]{}

    \newcommand{\abrm}[1]{}
    \newcommand{\abrcm}[1]{}

}

\begin{document}
\begin{frontmatter}
\title{MorteX method for contact along real and embedded surfaces:\\ coupling X-FEM with the Mortar method}
\author[cdm,st]{Basava R. Akula}
\ead{basava-raju.akula@mines-paristech.fr}
\author[st]{Julien Vignollet}
\ead{julien.vignollet@safrangroup.com}
\author[cdm]{Vladislav A. Yastrebov\corref{cor1}}
\ead{vladislav.yastrebov@mines-paristech.fr}
\cortext[cor1]{Corresponding author}
\address[cdm]{MINES ParisTech, PSL Research University, Center des Mat\'eriaux, CNRS UMR 7633, BP 87, 91003 Evry, France}
\address[st]{Safran Tech, Safran Group, 78772 Magny-les-Hameaux, France}
\begin{abstract}
A method to treat frictional contact problems along embedded surfaces in the finite element framework is developed.
Arbitrarily shaped embedded surfaces, cutting through finite element meshes, are handled by the X-FEM.
The frictional contact problem is solved using the monolithic augmented Lagrangian method within the mortar framework which was adapted for handling embedded surfaces.
We report that the resulting mixed formulation is prone to mesh locking in case of  high elastic and mesh density contrasts across the contact interface.
The mesh locking manifests itself in spurious stress oscillations in the vicinity of the contact interface. We demonstrate that in the classical patch test, these oscillations can be removed  simply by using triangular blending elements.
In a more general case, the triangulation is shown inefficient, therefore stabilization of the problem is achieved by adopting a recently proposed coarse-graining interpolation of Lagrange multipliers. 
Moreover, we demonstrate that the coarse-graining is also beneficial for the classical mortar method to avoid spurious oscillations for contact interfaces with high elastic contrast.
The performance of this novel method, called MorteX, is demonstrated on several examples which show as accurate treatment of frictional contact along embedded surfaces as the classical mortar method along boundary fitted surfaces. 

\end{abstract}


\end{frontmatter}

\section{Introduction} \label{sec:intro}

Among the wide spectrum of engineering applications, the class of problems
involving contact interactions between solids are complex both with regard to
their mathematical description and numerical treatment.
The transfer of mechanical forces, thermal or
electrical conduction, tectonic plate movements are merely a few examples of the
ubiquity of contact in nature and engineering applications. 
The vast majority of phenomena occurring at the contact interface, such as wear, adhesion,
lubrication and fretting, as well as mass and energy transfer, determine to the greater
extent the service life of engineering components. This lays a strong
emphasis on the importance to gain a fine understanding of these mechanisms for the accurate analysis
and timely prediction of failure.  In the last decade, a class of surface-to-surface contact discretizations, such as mortar method, coupled with appropriate treatment of inequality contact constraints has been well established and proved its ability to efficiently treat contact problems~\cite{puso_mortar_2004,fischer_mortar_2006,popp_mortar_2012}.
However, the numerical treatment of the contact and related phenomena remains
particularly challenging when it involves the evolution of complex surfaces. These
complexities can involve cumbersome remeshing algorithms and field transfer
procedures, as for instance those encountered in the context of wear. In addition,  construction of adequate finite element meshes near contact interfaces
and stability of the contact formulations are necessary ingredients to ensure the overall efficiency, accuracy and robustness of the numerical procedures.

\begin{figure}[htb!]
   \centering
   \includegraphics[width=.75\textwidth]{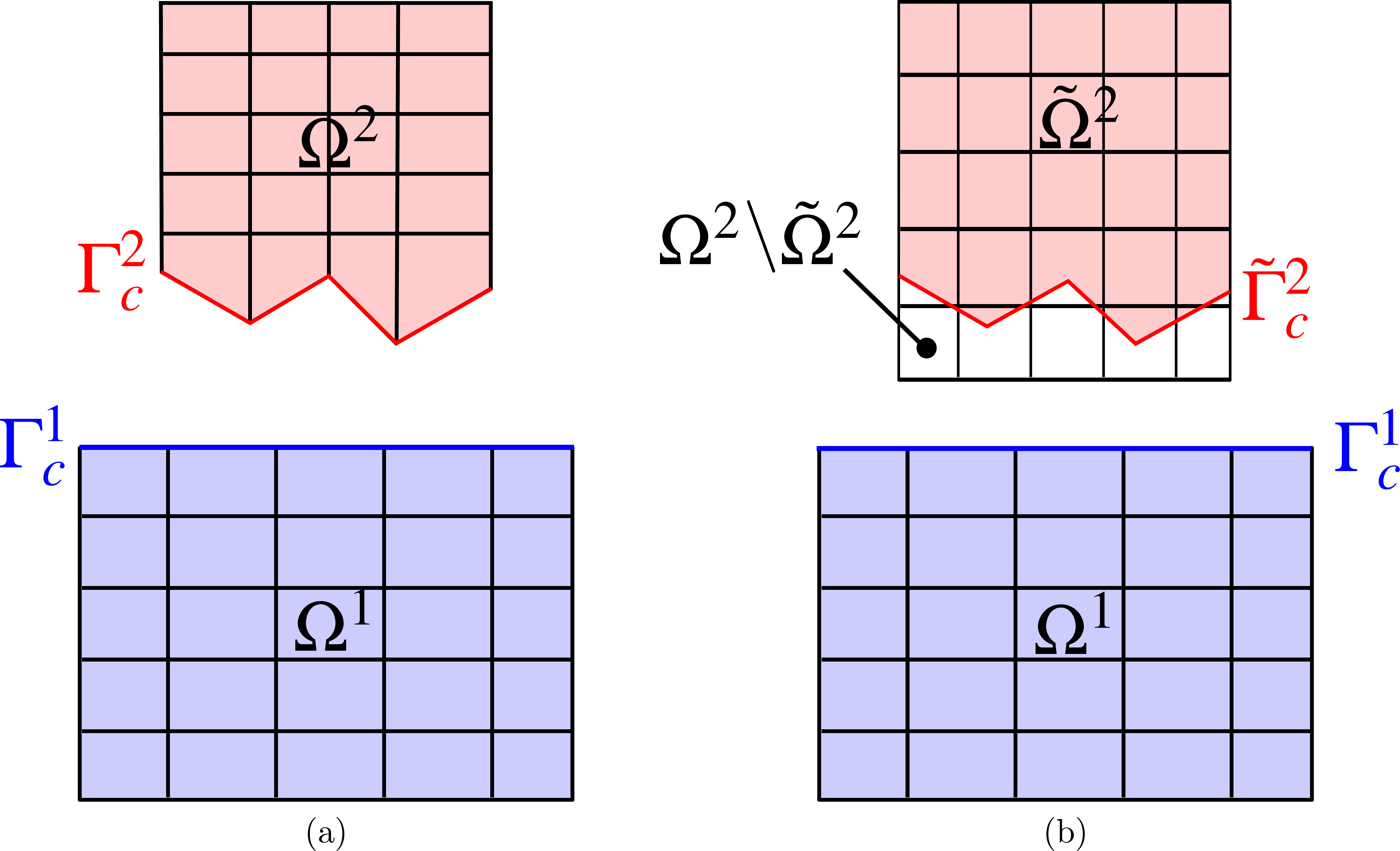}
   \caption{Equivalent discrete setting of two deformable bodies coming in contact: (a) the contact surfaces $\Gamma^1_c$ and $\Gamma^2_c$ are conformal to the boundaries of discretized solids; (b) contact boundary $\Gamma^1_c$ is still conformal, while the homologue deformable body is now $\tilde\Omega^2$ and its contact boundary $\tilde\Gamma^2_c$ is embeded in the domain $\Omega^2$, whose part $\Omega^2\setminus\tilde\Omega^2$ is discarded from the computation.}
   \label{fig:two_body_contact_Discrete}
\end{figure}

In this work, first we present a coherent framework for mortar frictional contact based on the monolithic augmented Lagrangian method for both normal and tangential contact components~\cite{alart_mixed_1991,cavalieri2015numerical}.
We then extend this framework for a two dimensional unified framework, called MorteX, capable to treat frictional contact problems between a real and a virtual (intra-mesh) surfaces. It  combines the features of Generalized/Extended Finite element methods (GFEM/X-FEM)~\cite{dolbow_extended_1999,belytschko_review_2009} and mortar methods for solving the contact problems~\cite{belgacem_mortar_1998,mcdevitt_mortar-finite_2000,puso_mortar_2004,popp_finite_2009}.  This unified framework was elaborated in~\cite{akula_tying_paper} for mesh tying problems and is extended here to handle frictional contact problems. 
Within the proposed framework, in a discretized setting (see Fig.~\ref{fig:two_body_contact_Discrete}) the contact constraints are imposed between a ``real'' surface $\Gamma_{c}^1$ (explicitly represented by edges of finite elements) and an embedded virtual surfaces $\Gamma^2_c$ (internal surface geometrically  non-conformal with the underlying discretization).

The X-FEM is an enrichment method based on the partition of unity (PUM) for discontinuous fields~\cite{melenk_partition_1996,babuska_partition_1997}.  In this framework, discontinuities are not explicitly modeled, and as a result, the mesh is not required to conform to these discontinuities.  
In X-FEM, enrichment functions are added to the finite element interpolation using the framework of PUM, to account for non-smooth behavior without compromising on the optimal convergence~\cite{ferte_interface_2014}. The X-FEM methods are extensively used in applications such as fracture mechanics, shock wave front and oxidation front propagation, and other applications involving discontinuities both strong and weak~\cite{sukumar_arbitrary_2000,sukumar_modeling_2001,diez_stable_2013,gross_extended_2007, ji_hybrid_2002}, and in particular for the contact between crack lips~\cite{dolbow_extended_2001,khoei_contact_2006,ribeaucourt_new_2007,elguedj_mixed_2007,liu_contact_2008,gravouil_stabilized_2011,mueller-hoeppe_crack_2012}. The ability of the X-FEM to describe geometric features (e.g. inclusions and voids) independently  of the underlying mesh,  will be exploited in the proposed unified framework to describe complex and evolving surfaces.

As opposed to the node-to-node discretisation which is limited to small sliding and the node-to-surface method which is inaccurate, the surface-to-surface based mortar methods are stable and very accurate in treating contact problems under finite sliding conditions~\cite{el_abbasi_stability_2001}.  
The mortar methods provide us with a comprehensive framework to address the limitations of incompatible interface discretizations.  
They are a subclass of domain decomposition methods (DDM), that are tailored for non-conformal spatial interface discretizations~\cite{bernardi_new_1994}, and were originally introduced for spectral elements~\cite{belgacem_spectral_1994,bernardi_coupling_1990}. 
The coupling and tying of different physical models, discretization schemes, and/or non-matching discretizations along interfaces between the domains can be ensured by mortar methods.  
The mathematical optimality and applicability of the mortar methods in spectral and finite element frameworks were studied extensively for elliptic problems in~\cite{bernardi_coupling_1990,belgacem_spectral_1994,wohlmuth_discretization_2001}. 
The mortar methods were brought into the contact realm with the initial contributions from Belgacem et al., Hild, Laursen et al. ~\cite{belgacem_mortar_1998,hild_numerical_2000,mcdevitt_mortar-finite_2000}. 
Subsequent works of Fischer and Wriggers ~\cite{fischer_frictionless_2005,fischer_mortar_2006}, Puso and Laursen ~\cite{puso_mortar_2004,puso_mortar_2004-1} have extended the application of mortar methods to the class of non-linear problems. 
The work of Popp, Gietterle, et al~\cite{gitterle_finite_2010,popp_mortar_2012} dealt with the class of dual Lagrangian resolution schemes ~\cite{wohlmuth_mortar_2000}.

In the present methodology, we use the monolithic augmented Lagrange multiplier resolution scheme to treat the frictional contact problems.
It represents a so-called mixed formulation~\cite{brezzi_mixed_2012}.  
The choice of Lagrange multipliers' functional space strongly affects the convergence rate and can lead to loss of accuracy in the interfacial tractions. 
These difficulties arise from a locking type phenomena reported for mixed variational formulations as a result of non-satisfaction of Ladyzhenskaya-Babu\v{s}ka-Brezzi (LBB)~\cite{babuska_finite_1973,brezzi_mixed_2012}. 
This locking, present in standard FEM with Lagrange multipliers on the boundary~\cite{barbosa_finite_1991}, but also in the mixed X-FEM framework when  Dirichlet boundary conditions are applied along embedded surfaces, was extensively studied in~\cite{fernandez-mendez_imposing_2004,moes_imposing_2006,bechet_stable_2009,burman_fictitious_2010,hautefeuille_robust_2012,ramos_new_2015}.
However, to the best of our knowledge,  the manifestation of mesh locking effects when imposing contact constraints using Lagrange multipliers, was never reported.
The manifestation of the locking effect will be illustrated in the present work for contact problems both in the context of standard mortar and MorteX methods. 
In addition, stabilization strategies initially proposed for the mesh tying problems in the context of MorteX method~\cite{akula_tying_paper}, are adapted here for contact applications.


The paper is organized as follows. 
In Section~\ref{sec:strong_form} and \ref{sec:weak_form}, we present the strong and weak continuum problem description of two-body frictional contact. 
The classical spatial mortar interface discretization scheme for the frictional contact problem is derived in Section~\ref{sec:mortar_interface_disc}.
In Section~\ref{sec:mortex_framework} this framework is extended to account for embedded contact surfaces, and referred as MorteX framework. It will explicitly detail the features of the X-FEM and adaptation of the mortar which are needed to construct a coherent computational scheme.
In Section~\ref{sec:examples}, few examples including the contact patch test are considered to illustrate the methods performance for both frictionless and frictional contact problem settings. 
In Section~\ref{sec:conclusion}, we give the concluding remarks and present the prospective works.

\section{Problem definition\label{sec:strong_form}}

\begin{figure}[htb!]
   \centering
   \includegraphics[width=1\textwidth]{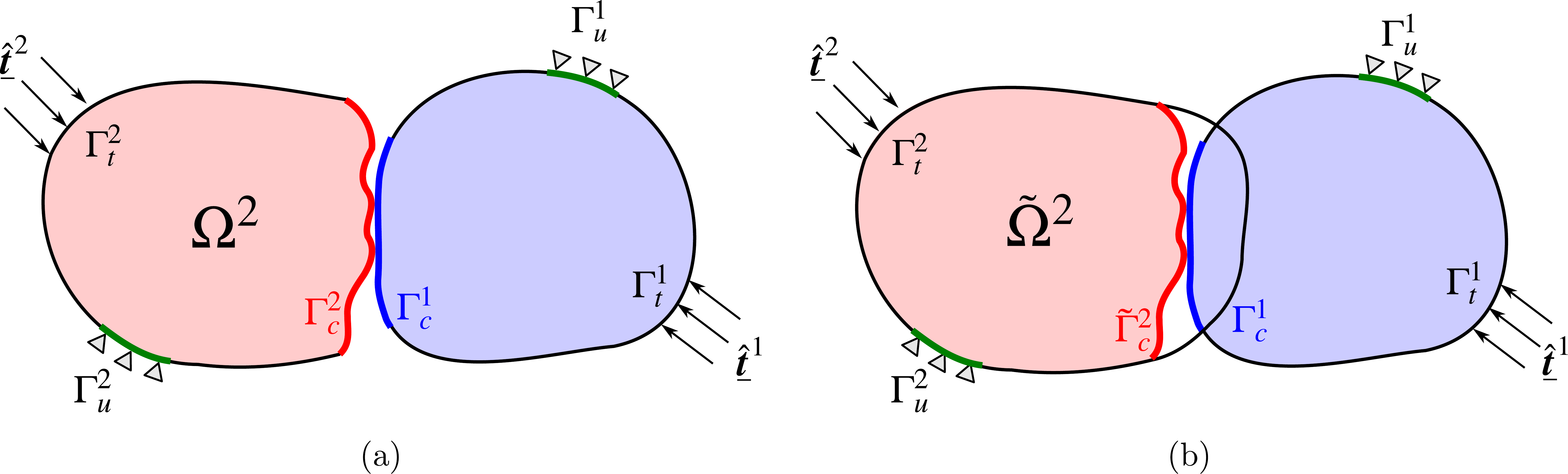}
   \caption{Continuum setting of two deformable bodies $\Omega^1$ and $\Omega^2$
   coming into contact.}
   \label{fig:two_body_contact_Continuum}
\end{figure}

The case of two deformable bodies in contact is considered, for simplicity the derivations are made in  infinitesimal deformation formalism, but is readily extandable to the finite strain framework. 
Fig.~\ref{fig:two_body_contact_Continuum} shows two statements of this contact problem: (a) is a classical formulation where the solids come in contact over their outer surfaces. In (b) an equivalent, from continuum point of view, setting is shown, whereby
contact boundary $\Gamma^1_c$ is still an outer boundary of $\Omega^1$, while the homologue deformable body is now $\tilde\Omega^2$ and its contact boundary $\tilde\Gamma^2_c$ is embeded in the domain $\Omega^2$, whose part $\Omega^2\setminus\tilde\Omega^2$ is discarded from the computation.
In the current and next two sections, we will state the problem and suggest the methods to handle the configuration shown in Fig.~\ref{fig:two_body_contact_Continuum}(a), whereas in Section~\ref{sec:mortex_framework}, we will extend these methods to deal with the problem shown in Fig.~\ref{fig:two_body_contact_Continuum}(b).
The domains $\Omega^{i}$, $i=1,\,2$, have non-intersecting Dirichlet boundaries $\Gamma_{u}^i$, Neumann boundaries $\Gamma_{t}^i$ and contact boundaries $\Gamma_{c}^i$, i.e. $\partial{\Omega^{i}}=\Gamma_{u}^{i} \cup \Gamma_{t}^{i}\cup\Gamma_{c}^{i}$. 
The standard boundary value problem (BVP) is described by the balance of linear momentum along with the imposed boundary conditions:
\begin{align}
    \nabla\cdot\ten{\sigma}^i+\vec{f}_v^i &= 0\quad \mbox{in}\,\,\Omega^i,
    \label{eq:strong_form_linear_momentum}\\
    \ten{\sigma}^i\cdot\vec{n}^i &=
    \vec{\hat{t}}^i\quad\mbox{on}\,\,\Gamma_t^i,\label{eq:strong_form_NeumannBC}\\
    \vec{u}^i &=
    \vec{\hat{u}}^i\quad\mbox{on}\,\,\Gamma_u^i,\label{eq:strong_form_DirichletBC}\\
    \ten{\sigma}^i\cdot\vec{n}^i &=
    \vec{t}^i_c\quad\mbox{on}\,\,\Gamma_c^i,\quad i=1,2.\label{eq:strong_form_Contact}
\end{align}
$\ten{\sigma}^i$ is the Cauchy stress tensor. Under the small deformation assumption the strain-displacement and constitutive relations are:  
\begin{equation} 
  \ten\varepsilon=\frac{1}{2}\big[(\nabla\vec u)^\intercal+\nabla\vec u\big],\quad \ten\sigma=\ten{\mathcal{C}}:\ten\varepsilon \label{eq:hookes_law} 
\end{equation}
where $\ten\varepsilon$ and $\ten{\mathcal{C}}$ are the Cauchy strain tensor  and the fourth-order elasticity tensor describing Hooke's  law, respectively. $\vec{f}_v^i$ represents the density of body forces, $\vec{n}^i$ is the unit outward normal to $\Omega^i$ and  $\vec{\hat t}^i$, $\vec{\hat{u}}^i$ are the prescribed tractions and displacements on $\Gamma_t^i$ and $\Gamma_u^i$, respectively. 
The contact between the bodies introduces constraints which can be treated as additional configuration-dependent traction ($\vec{t}_c^i$), which have to be only compressive (in absence of adhesion) and ensure non-penetration of solids~\eqref{eq:strong_form_Contact}.

The gap function defines the signed distance between the two surfaces and is the fundamental measure that determines the status of contact. 
Here, we define the gap vector as a vector from a point $\vec x^1$ on the surface $\Gamma_c^1$ to its closest fprojection $\hat{\vec x}^2(\vec x^1)$ along the normal $\vec n(\vec x^1)$ onto the surface $\Gamma_c^2$ (the choice of these surfaces is arbitrary):
\begin{equation}    
    \vec g(\vec x^1,\Gamma_c^2) =   \hat{\vec x}^2(\vec x^1)  - \vec x^1   \label{eq:gap_vector_continuous}.
\end{equation}
The normal gap function is thus obtained as a dot product of the gap vector $\vec g$~\eqref{eq:gap_vector_continuous} and the outward unit normal $\vec n(\vec x^1)$ 
\begin{equation}     
    g_n(\vec x^1,\Gamma_c^2) = \vec g\cdot\vec n(\vec x^1).     \label{eq:normal_gap} 
\end{equation}

The MorteX method is, at this stage, elaborated for two-dimensional problems only, which allows some simplifications in case of frictional contact, which is also presented in 2D. 
The tangential relative sliding velocity is given as in~\cite{puso_mortar_2004}:
\begin{equation}     
\dot g_\tau(\vec x^1) = \vec \tau(\vec x^1)\cdot\left(\dot{\vec x}^1  - \dot{\hat{\vec x}}^2(\vec x^1)\right)     \label{eq:tang_rel_vel} 
\end{equation} 
where $\vec\tau(\vec x^1)$ is the unit tangential vector to $\Gamma_c^1$ at point $\vec x^1$.
In order to preserve the consistency of units, hereinafter the incremental slip will be used instead of the slip velocity $\mathring g_\tau = \dot g_\tau \Delta t$, where $0 < \Delta t < \infty$ is an arbitrary time increment. This replacement does not change the physical sense of all following equations.
With these notations at hand, the contact traction can be decomposed into normal and tangential contributions: 
\begin{equation}     
\vec t_c^1 = p_n\vec n+p_\tau\vec\tau.
\label{eq:contact_trca_decompose}
\end{equation} 
Along the contact interface, the classical Hertz-Signorini-Moreau conditions~\cite{kikuchi_contact_1988} also known as KKT conditions in the theory of  optimization~\cite{hestenes_multiplier_1969,powell_algorithms_1978-1} have to be satisfied. These conditions are formulated using the normal gap $g_n$ and the contact pressure $p_n$ as:
\begin{equation}     
g_n\geq 0,\quad p_n\leq 0,\quad p_ng_n=0.     \label{eq:gn_string_form}
\end{equation}   
The tangential friction is described using the Coulomb friction law.
The frictional constraints are formulated using the tangential slip increment $\mathring g_\tau$, the tangential traction $p_\tau$ and the contact pressure $p_n$.  
The incremental slip vanishes $|\mathring g_\tau|=0$ when the tangential traction is below the frictional threshold $|p_\tau| \le \mu|p_n|$, non-zero incremental slip  $|\mathring g_\tau|\ne0$ is possible when $|p_\tau| =\mu|p_n|$. These conditions can be also formulated as KKT conditions:   
\begin{equation}  
|\mathring g_\tau|\ge0,\quad|p_\tau|-\mu|p_n|\le0, \quad  \left(\mu|p_n|-|p_\tau|\right)
|\mathring g_\tau|=0  \label{eq:gt_string_form}
\end{equation}
where $\mu$ is the Coulomb's coefficient of friction.

\section{Weak form\label{sec:weak_form}}
The Eqs.~\eqref{eq:strong_form_linear_momentum}-\eqref{eq:strong_form_DirichletBC} denote the strong form of the standard solid mechanics BVP, and Eq.~\eqref{eq:strong_form_Contact}, along with the contact conditions~\eqref{eq:gn_string_form}-\eqref{eq:gt_string_form} represents the contact part of the BVP.  
The weak form of the contact problem reduces to a variation inequality and the associated displacements should be selected in a constrained functional space, which has to include the non-penetration condition $g_n\ge0$~\cite{fichera_boundary_1973,kikuchi_contact_1988}.
It is a constrained minimization problem. However, using the augmented Lagrangian method permits us to convert this problem into a fully unconstrained one, with displacements $\vec u$ and virtual displacements $\delta\vec u$ (test functions) selected in unconstrained functional spaces:
$\mathcal{U}^i$ and the test function space $\mathcal{V}^i$ definitions: 
\begin{align} 
    \mathcal{U}^i &= \{u^i\in H^1(\Omega^i)\,|\,u^i=\hat{u}^i\,\mbox{on}\,\Gamma_u^i\},\\ \mathcal{V}^i &= \{\delta u^i\in H^1(\Omega^i)\,|\,\delta u^i=0\,\mbox{on}\,\Gamma_u^i\} \label{eq:weak_form_structural} 
\end{align} 
where $H^1(\Omega^i)$ denotes the first order Sobolev space. The virtual work $\delta W_s$ corresponding to the structural part is given by:  
\begin{align}     
    \delta W_s =     \sum_{i=1}^{2}\bigg[\underbrace{\int\limits_{\Omega^i}\ten{\sigma}^i:\delta\ten\varepsilon^i\,d\Omega^i}_{\delta W_{int}^i}-\underbrace{\int\limits_{\Omega^i}\vec{f}_v^i\cdot\delta\vec     u^i\,d\Omega^i-\int\limits_{\Gamma_t^i}\hat{\vec t}^i\cdot\delta\vec     u^i\,d\Gamma_t^i}_{\delta W_{ext}^i}\bigg],  \label{eq:virtual_work_structural} 
\end{align}
where $\delta W_{\text{int}},\delta W_{\text{ext}}$ are the change in internal energy and the virtual work of forces, respectively.

Regarding the contribution of the virtual work associated with the contact problem, we adapt the monolithic augmented Lagrangian method (ALM) framework of Alart and Curnier~\cite{alart_mixed_1991} to the mortar framework.
The ALM is a mixed dual formulation that introduces Lagrange multipliers as dual variables along with the primal displacement variables. 
The fields of Lagrange multipliers $\lambda_n$ and $\lambda_\tau$ are respectively equivalent to the contact pressure $p_n$ and the tangential friction shear $p_\tau$ introduced in Eq.~\eqref{eq:contact_trca_decompose}.
They and their virtual variations are selected from the appropriate functional space $\lambda_{n,\tau},\delta\lambda_{n,\tau} \in H^{-1/2}$, namely fractional Sobolev space of order $-1/2$.
The augmented Lagrangian functionals $l_n$ and $l_\tau$ for the normal and frictional contact respectively are given by the following expressions~\cite{pietrzak_continuum_1997,pietrzak_large_1999}: 
\begin{equation}     
l_n(g_n,\lambda_n)=    \begin{cases} \lambda_n g_n+\frac{\varepsilon_n}{2}g_n^2,\quad&\,\hat{\!\lambda}_n\leq 0, \\    -\frac{1}{2\varepsilon_n}\lambda_n^2,\quad&\,\hat{\!\lambda}_n>0    \end{cases} \label{eq:ln} 
\end{equation} 
\begin{equation}   
l_\tau(\mathring g_\tau,\lambda_\tau,\,\hat p_n)= \begin{cases}     \begin{Bmatrix}     &\lambda_\tau     \mathring g_\tau+\frac{\varepsilon_\tau}{2}\mathring g_\tau^2,&\quad|\,\hat{\!\lambda}_\tau|\leq-\mu\,\hat p_n\\     
&-\frac{1}{2\varepsilon_\tau}\left(\lambda_\tau^2 +2\mu \hat p_n|\,\hat{\!\lambda}_\tau|+\mu^2 \hat p_n^2\right),&\quad|\,\hat{\!\lambda}_\tau|>-\mu\,\hat p_n     \end{Bmatrix},&\,\hat{\!\lambda}_n\leq0\\
  -\frac{1}{2\varepsilon_\tau}\lambda_\tau^2,&\,\hat{\!\lambda}_n>0 \end{cases} \label{eq:lt} 
\end{equation} 
where $\,\hat{\!\lambda}_n,\hat{\!\lambda}_\tau,\hat p_n$ are the augmented normal and tangential Lagrange multipliers, and the augmented pressure, respectively:
\begin{equation} 
\,\hat{\!\lambda}_n=\lambda_n+\varepsilon_n g_n,\quad \,\hat{\!\lambda}_\tau=\lambda_\tau+\varepsilon_\tau \mathring g_\tau,\quad \hat p_n = p_n+\varepsilon_n g_n \label{eq:aug_lags},
\end{equation} 
where $\varepsilon_n$ and $\varepsilon_\tau$ are the normal and tangential augmentation parameters, respectively.
The augmented pressure determining the frictional threshold is replaced by the corresponding augmented Lagrange multiplier $\hat{\!\lambda}_n$, however $\hat p_n$ is not subjected to variation, therefore a different notation is used to highlight this subtle difference.
Integrating the functionals~\eqref{eq:ln} and \eqref{eq:lt} over the contact surface $\Gamma_c^1$ integrates the contact constraints in the Lagrangian: 
\begin{equation}   
\mathcal L = W_s + W_c=W_s+\int\limits_{\Gamma_c^1}l_n(g_n,\lambda_n)+ l_\tau(\mathring g_\tau,\lambda_\tau,\hat p_n)\,d\Gamma, \label{eq:continuos_contact_energy} 
\end{equation}
where $W_s$ is the potential energy of the system.

The  problem of contrained optimization  transforms into the min-max or saddle point problem for the Lagrangian, whose solution is equivalent to the solution of the variational inequality optimization problem on a non-convex domain and its boundary~\cite{hestenes_multiplier_1969,powell_algorithms_1978-1,pietrzak_continuum_1997}. The min-max problem is solved by searching the stationary point minimizing the Lagrangian with respect to primal (displacement) field and maximizing it with respect to the dual field (Lagrange multipliers). The variation of the structural part, which contains only primal variables is given in~\eqref{eq:virtual_work_structural}, whereas the variation of the contact contribution is given by:
\begin{align} 
\delta W_c&=\int\limits_{\Gamma_c^1}\delta l_n(g_n,\lambda_n)+\delta l_\tau(\mathring g_\tau,\lambda_\tau,\hat p_n)\,d\Gamma_c^1\label{eq:tot_cont_vw} = \\           
&=\int\limits_{\Gamma_c^1}\frac{\partial l_n}{\partial g_n}\delta g_n+\frac{\partial           l_\tau}{\partial \mathring g_\tau}\delta           \mathring g_\tau\,d\Gamma_c^1\,+&\quad\mbox{frictional contact virtual           work}\label{eq:contact_VW}\\     
&\quad+\int\limits_{\Gamma_c^1}\frac{\partial l_n}{\partial \lambda_n}\delta           \lambda_n\,d\Gamma_c^1\,+&\quad\mbox{weak normal contact           contribution}\label{eq:weak_normal_condition}\\   
&\quad+\int\limits_{\Gamma_c^1}\frac{\partial l_\tau}{\partial \lambda_\tau}\delta           \lambda_\tau\,d\Gamma_c^1&\quad\mbox{weak tangential contact           contribution}.\label{eq:weak_tang_condition} 
\end{align}

The contact virtual work~\eqref{eq:tot_cont_vw} is expressed as a summation of the contact contribution to the virtual work resulting from the variations of the positions~\eqref{eq:contact_VW}, the weak contribution of normal contact constraints~\eqref{eq:weak_normal_condition} from the variation of the normal Lagrange multiplier and the weak contribution of tangential contact constraints~\eqref{eq:weak_tang_condition} from the variation of the tangential Lagrange multiplier. Note that the variations and derivatives of piece-wise smooth potentials $l_n$ and $l_\tau$ can be seen as sub-derivatives~\cite{pietrzak_continuum_1997}. We recall that the variation of $\hat p_n$ is not required. Based on the three possible contact statuses (i.e. stick, slip or no contact), the contact surface is divided into three distinct sub-surfaces: 
\begin{equation*}  
\Gamma_c^1 = \Gamma_{{\mbox{\tiny stick}}}\,\cup\,\Gamma_{{\mbox{\tiny  slip}}}\,\cup\,\Gamma_{{\mbox{\tiny nc}}}. 
\end{equation*}
The frictional contact contribution to the virtual work can be split as follows:   
\begin{equation}  
\delta W_c =   
\begin{cases}\displaystyle \; 
  \int\limits_{\Gamma_{{\mbox{\tiny stick}}}}\,\hat{\!\lambda}_n\delta  g_n+g_n\delta\lambda_n+\,\hat{\!\lambda}_\tau\delta\mathring g_\tau+\mathring g_\tau\delta \lambda_\tau\,d\Gamma,&{\hat{\!\lambda}_n\le0,\;}|\,\hat{\!\lambda}_\tau|\leq-\mu\,\hat p_n,\mbox{ stick,}\\ 
  \displaystyle   \;\int\limits_{\Gamma_{{\mbox{\tiny slip}}}}\,\hat{\!\lambda}_n\delta  g_n+g_n\delta\lambda_n-\mu\,\hat p_n\mathrm{sign}(\,\hat{\!\lambda}_\tau) \delta\mathring g_\tau-\frac{1}{\varepsilon_\tau}\bigg(\lambda_\tau+\mu\,\hat{p}_n \mathrm{sign}(\,\hat{\!\lambda}_\tau)\bigg)\;\delta \lambda_\tau\,d\Gamma,&{\hat{\!\lambda}_n\le0,\;}|\,\hat{\!\lambda}_\tau|>-\mu\,\hat{p}_n,\mbox{slip,}\\ \displaystyle \;\int\limits_{\Gamma_{{\mbox{\tiny nc}}}}-\frac{1}{\varepsilon_n}\lambda_n\delta\lambda_n-\frac{1}{\varepsilon_\tau}\lambda_\tau\delta\lambda_\tau{\,d\Gamma},&\,\hat{\!\lambda}_n>0.  
  \end{cases}  \label{eq:vw_final} 
\end{equation}

\section{Mortar interface discretization}\label{sec:mortar_interface_disc}

Within the mortar discretized framework, the contacting surfaces are typically classified into mortar and non-mortar sides. 
In the following, the superscript ``$1$'' refers to the mortar side of the interface and ``$2$'' to the non-mortar side.  The choice of mortar and non-mortar is rather arbitrary: however, in general the choice of the finer meshed surface as mortar side is known to result in higher accuracy. 
All the contact related integral evaluations will be carried out on the mortar side of the interface. 
In the considered two dimensional case, the edges of potentially contacting surfaces are parametrized by $\xi^i\,\in[-1;1]$ .  
In the current configuration the interpolations
of mortar and non-mortar surface edges are given by: 
\begin{equation} \vec x^1(\xi^1,t) = N^1_m(\xi^1)\vec x_m^1,\quad {\vec x^2(\xi^2,t) = N^2_i(\xi^2)\vec x_i^2, \quad m\in[1,\mathrm{M}],\;i\in[1,\mathrm{N}]}. \label{eq:geom_interpolations} 
\end{equation}
$N^1_m(\xi^1)$ and $N^2_i(\xi^2)$ are the interpolation functions of mortar and non-mortar sides, respectively; whereas $\mathrm{M}$ and $\mathrm{N}$ is the number of nodes per segment at mortar and non-mortar sides, respectively.  
Note that we use isoparametric linear elements, which implies that the normal remains constant along the edge.
Moreover, that in the current framework, for the sake of simplicity, we omitted averaging of normals at nodes and their interpolation, which was elaborated in~\cite{yang_two_2005,popp_mortar_2012}. 
Hereinafter, Einstein summation over repetitive index is used.
We assign  scalar Lagrange multipliers $\lambda_n$ and $\lambda_\tau$ for the normal and tangential directions to the nodes of the mortar side. The Lagrange multiplier vector $\vec\lambda = \lambda_n\vec n +\lambda_\tau\vec \tau$ is interpolated functions $\Phi_l$:
\begin{equation} 
{ \vec\lambda(\xi^1,t) = \Phi_l(\xi^1)\vec\lambda^l(t),\quad l\in[1,\mathrm{L}] } \label{eq:lambda_interpolations} 
\end{equation} 
where $\mathrm{L}$ is the number of nodes used for interpolation over every edge, where L can be less than or equal to M.

\begin{figure}[htb!] 
\centering \includegraphics[width=.75\textwidth]{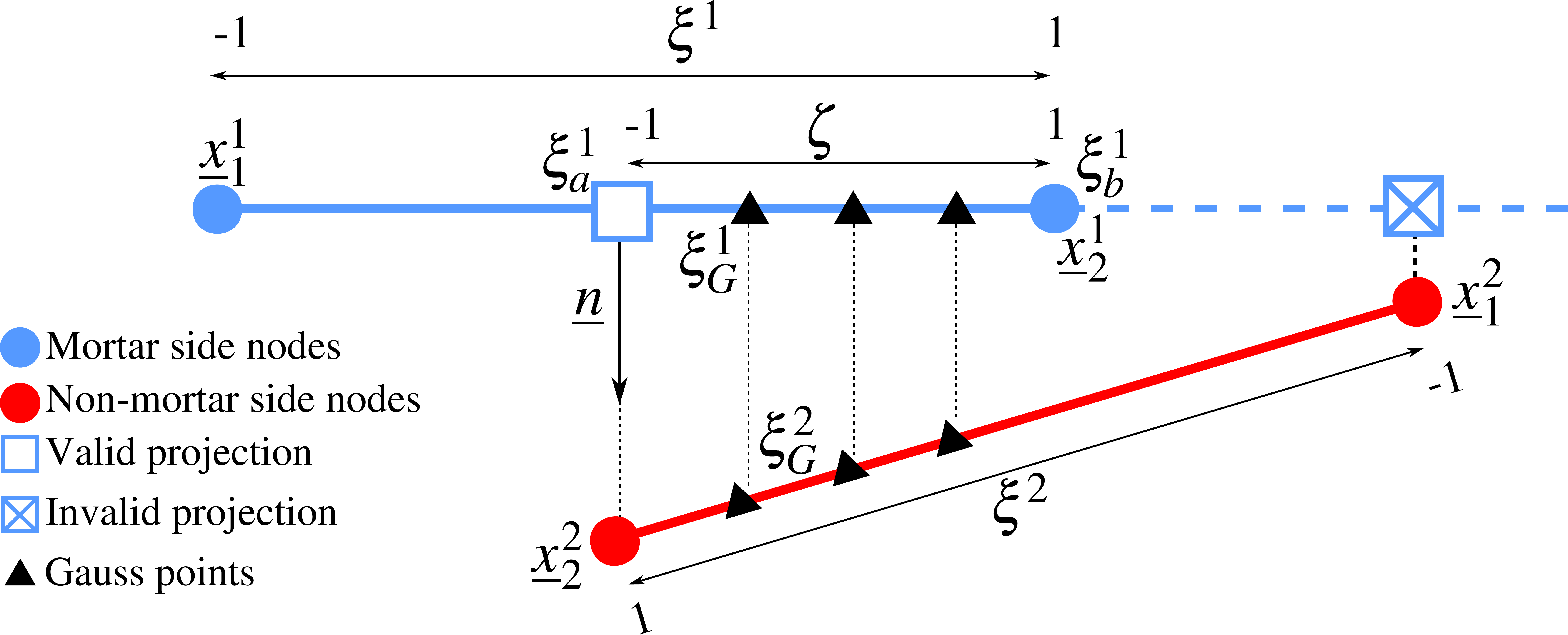} 
\caption{     Nodes of the non-mortar edge are     projected onto mortar edge, establishing a mapping and determining the integration domain parametrized by $\zeta$. }
\label{fig:mortar_integ_eval} 
\end{figure}

Every \emph{mortar contact element} is created between parts of mortar and non-mortar segments which are projected one on the other (see Fig.~\ref{fig:mortar_integ_eval}). The segment-to-segment projection is described in detail in~\cite{wriggers_computational_2006,yang_two_2005,popp_finite_2009} and will not be reproduced here.  
Each contact element consists of $\mathrm{M+N}$ nodes, each with $2$ primal DoFs in 2D, and of $\mathrm{L}$ or $2\mathrm{L}$ dual DoFs associated with Lagrange multipliers in case of frictionless or frictional contact, respectively.  

\subsection{Discrete integral kinematic quantities}\label{subsec:disc_integ_qtys}

As a first step towards the elaboration of the contact part of the residual vector, the discretisations~\eqref{eq:geom_interpolations} and~\eqref{eq:lambda_interpolations} are inserted in the continuous weak form~\eqref{eq:vw_final}. First, we focus on the derivation of a single term of this integral, which enables us to introduce coincise notations and simplifications to be used for the other terms.
It is  obtained for every active contact element by integrating over a  part or an entire mortar edge $S^{\mathrm{el}}$: 
\begin{equation}
\int\limits_{S^{\mathrm{el}}}g_n\delta \lambda_n\,d\Gamma = \int\limits_{S^{\mathrm{el}}} \left(N^1_m(\xi^1)\vec x^1_m - N^2_i(\xi^2)\vec x^2_i\right)\cdot\vec n\,\Phi_l(\xi^1)\delta\lambda^l_n\,d\Gamma \label{eq:integral_gn_derivation}
\end{equation} 
where $\vec n$ is the unit outward normal of the mortar edge. Note that this equality holds if the parametrization $\xi^2$ ensures collinearity between the vector in brackets and the normal vector. Therefore, $\xi^2$ depends on positions of all the nodes both mortar and non-mortar forming the contact element. This is not explicitly reflected in the equation but should be kept in mind. The hat-notations, which denoted projections in~(\ref{eq:gap_vector_continuous},\ref{eq:tang_rel_vel}), are therefore omitted hereinafter.

It appears useful~\cite{puso_mortar_2004} to introduce an integral
gap vector $\bar{\! \vec g}^{l}$, which is defined using the above 
Eq.~\eqref{eq:integral_gn_derivation} as:
\begin{equation}
{\bar{\! \vec g}^{l}(t)}= \left[{D}_{lm}\vec x_m^1 - {M}_{li}\vec x_i^2\right]
\label{eq:integral_gap_vector}
\end{equation}
where  $D_{lm}$ and  $M_{li}$ are the so-called mortar
integrals~\cite{popp_finite_2009}, which are defined as\\
\begin{tabularx}{\textwidth}{XX}
\begin{equation}
     D_{lm} = \int\limits_{S^{\mathrm{el}}} \Phi_l(\xi^1) N^1_m(\xi^1)\,d\Gamma,\quad
    \label{eq:mortar_integrals_D}
\end{equation}&
\begin{equation}
     M_{li} =  \int\limits_{S^{\mathrm{el}}}\Phi_l(\xi^1)
    N^2_i(\xi^2)\,d\Gamma.
    \label{eq:mortar_integrals_M}
\end{equation}
\end{tabularx}
The evaluation of the these integrals forms the core of the mortar
discretization scheme. The evaluation of these integrals is detailed in
Section~\ref{subsec:mort_integ_eval}. 
Using the introduced notations, the term~\eqref{eq:integral_gn_derivation} takes a simple form:
\begin{equation}
   \int\limits_{S^{\mathrm{el}}}g_n\delta \lambda_n\,d\Gamma = \bar{\!g}_{n}^l\delta\lambda^l_n
\end{equation}
where 
\begin{equation}
 \bar g_n^l = \bar{\vec g}^l\cdot \vec n = \left[D_{lm}\vec x^1_m - M_{li}\vec x^2_i\right]\cdot\vec n
 \label{eq:integgn}
\end{equation}
%
is the integral normal
gap. The variation of the integral normal gap is given by:
\begin{equation}
\delta\bar{\! g}_{n}^l = \left[{D}_{lm}\delta \vec x_m^1 -
{M}_{li}\delta \vec x_i^2\right]\cdot\vec n.
\label{eq:gn_var}
\end{equation}
By analogy, the contribution of the tangential contact term can be introduced as:
 \begin{equation}
\int\limits_{S^{\mathrm{el}}}\mathring g_\tau\delta \lambda_\tau\,d\Gamma =
\mathring{\bar{\! g}}_{\tau}^l \delta\lambda^l_\tau,
\label{eq:integral_gt_derivation} 
 \end{equation}
where $\mathring{\bar g}_\tau^l$ is the integral slip increment given by~\cite{puso_mortar_2004}:
 \begin{equation}
 \begin{split}
   \mathring{\bar{\! g}}_{\tau}^l(t_j) &=
   -\vec\tau_l(t_j) \cdot\bigg( [D_{lm}(t_j)-D_{lm}(t_{j-1})]\vec x_m^1(t_{j-1}) -
    [M_{li}(t_j)-M_{li}(t_{j-1})]\vec x_i^2(t_{j-1})\bigg) =\\
    &= 
-\vec\tau_l(t_j) \cdot\left(D_{lm}(t_j)\vec x_m^1(t_{j-1}) - M_{li}(t_j)\vec x_i^2(t_{j-1}) - \vec g^l(t_{j-1})\right)
\end{split}
\label{eq:gt_var}
 \end{equation}
where the current time step $t_j$ and the previous time step $t_{j-1}$ come in
an interplay. The variation of the incremental slip, again following~\cite{puso_mortar_2004} is given by:
\begin{equation}
\delta \mathring{\bar{g}}_{\tau}^l(t_j) = \vec\tau_l(t_j) \cdot\bigg( D_{lm}(t_j)\delta\vec x_m^1-
    M_{li}(t_j)\delta\vec x_i^2\bigg).
    \label{eq:var_integral_increm_slip}
\end{equation}
Eqs.~\eqref{eq:gn_var} and \eqref{eq:var_integral_increm_slip} are both used for evaluation of Eq.~\eqref{eq:vw_final}.

\subsection{Evaluation of mortar integrals}\label{subsec:mort_integ_eval}

The evaluation of the mortar side integrals $D_{lm}$ is straightforward, as
it only involves the product of shape functions from the mortar
side~\eqref{eq:mortar_integrals_D}. This is in contrast with the evaluation of the mortar integral $M_{li}$ which combines shape functions from both the mortar and non-mortar sides. Its evaluation requires a mapping
between them (Fig.~\ref{fig:mortar_integ_eval}). For
this purpose, non-mortar quantities are projected onto the mortar side of the
interface using the mortar segment normal vector.   
The non-mortar nodes $\vec x_i^2\,(i\,\in[1,N])$ are projected onto the
mortar side segment along the mortar segment normal $\vec n$. The local coordinates $\xi^{1}_i$ of the projections are found by solving
\begin{equation} \label{eq:classical_xi_proj}
    \left(N^1_m(\xi_i^1)\vec x^1_m - \vec x_i^2\right)\times \vec n = 0.
\end{equation}
The extremities of the integration domain $S^{\mathrm{el}}$ are defined either by mortar or non-mortar edge nodes, details of this procedure are provides in~\cite{yang_two_2005} and not repeated here.
The segment of the mortar edge over which the evaluation of mortar integrals is carried out is parametrized by $\zeta\in[-1;1]$. The projection coordinates 
$\xi_{a/b}^{1}$, which determine the limits of the integration domain, are mapped on
the segment parameterization:
\begin{equation} \label{eq:classical_map_xi_eta}
    \xi^1(\zeta) =
    \frac{1}{2}(1-\zeta)\xi_{a}^{1}\,+\,\frac{1}{2}(1+\zeta)\xi_{b}^{1}.
\end{equation}
Fig.~\ref{fig:mortar_integ_eval} illustrates how the integration domain is determined between two edges. To evaluate the integrals using Gauss quadratures, the mortar-side Gauss points
$\xi^1_G$ are projected along mortar segment normal $\vec n$
onto the non-mortar side and the
corresponding local coordinates $\zeta_G$ are determined by:
\begin{equation}
    \left[N^2_i(\xi^2_G)\vec x^2_i - N^1_m(\xi^1_G)\vec x^1_m\right]\times \vec n = 0.
\end{equation}
The mortar matrices, evaluated with Gauss quadratures, take the following form:
\begin{equation}
D_{lm} =
    \int\limits_{S^{\mathrm{el}}}\Phi_l(\xi^1)N^1_m(\xi^1)\,d\Gamma=
    w_{G}\Phi_{l}(\xi^{1}_{G})N^1_{m}(\xi^{1}_{G})J_{\mbox{\tiny
    seg}}(\xi_{G}^{1}),
    \label{eq:mortar_D_op}
\end{equation}
\begin{equation}
 M_{li} =
    \int\limits_{S^{\mathrm{el}}}\Phi_l(\xi^1)N^2_i(\xi^2)\,d\Gamma
    = w_{G}\Phi_{l}(\xi^{1}_{G})N^2_{i}(\xi^{2}_{G})J_{\mbox{\tiny
    seg}}(\xi_{G}^{1}),
    \label{eq:mortar_M_op}
\end{equation}
where, as previously, $l\in [1,L]$, $m\in[1,M]$, $i\in[1,N]$ and $G \in [1,N_G]$, where $N_G$ is number of Gauss integration points, $J_{\mbox{\tiny seg}}$ is the
normalized Jacobian 
\begin{equation}
    J_{\mbox{\tiny seg}}(\xi_{G}^{1}) =
    \bigg|\frac{\partial{N^1_i}}{\partial\xi^{1}} \frac{\partial \xi^1}{\partial \zeta}\vec
    X_{i} \bigg|.
    \label{eq:2D_jacob}
\end{equation}
The factor $\partial \xi^1/\partial \zeta$ 
 reflects the fact that the integral is evaluated only over a part of the mortar edge.

\subsection{Residual vector}

Within the augmented Lagrangian formulation, every created contact element contributes to the virtual work of the system irrespectively of its contact status (active or inactive).
This translates into a smoother energy potential in comparison to the standard method of Lagrange multipliers.
For each contact element, the discretized form of the virtual work can be
obtained for the three possible contact statuses, namely stick, slip and no contact as follows:

%

\begin{align}
\delta W_{\mbox{\tiny stick}}^e = &
\begin{bmatrix}
 \delta\vec x_m^1\\[3pt]
 \delta\vec x_i^2\\[3pt]
 \delta\lambda_{n}^l\\[3pt]
 \delta\lambda_{\tau}^l\\
\end{bmatrix}^\intercal
\begin{bmatrix}
 D_{lm}\bigg(\hat{\!\lambda}_{n}^l\vec
 n+\hat{\!\lambda}_{\tau}^l\vec\tau\bigg)\\
 - M_{li}\bigg(\hat{\!\lambda}_{n}^l\vec
 n+\hat{\!\lambda}_{\tau}^l\vec\tau\bigg)\\
 \bar{g}_{n}^l\\[3pt]
 \mathring{\bar{g}}_{\tau}^l\\
\end{bmatrix},&\quad\hat{\!\lambda}_{n}^l\leq0,\,|\,\hat{\!\lambda}_{\tau}^l|\leq-\mu\,\hat{\!\lambda}_{n}^l\,(\mbox{stick})\label{eq:mortar_stick_resi}\\
\delta W_{\mbox{\tiny slip}}^e = &
\begin{bmatrix}
 \delta\vec x_m^1\\[3pt]
 \delta\vec x_i^2\\[3pt]
 \delta\lambda_{n}^l\\[3pt]
 \delta\lambda_{\tau}^l\\
\end{bmatrix}^\intercal
\begin{bmatrix}
 D_{lm}\bigg(\hat{\!\lambda}_{n}^l\vec
n-\mu\,\hat{\!\lambda}_{n}^l    \mathrm{sign}(\,\hat{\!\lambda}_{\tau})\vec\tau_l\bigg)\\
 -M_{li}\bigg(\hat{\!\lambda}_{n}^l\vec n-\mu\,\hat{\!\lambda}_{n}^l\mathrm{sign}(\,\hat{\!\lambda}_{\tau})\vec\tau_l\bigg)\\
 \bar{g}_{n}^l\\
-\frac{1}{\varepsilon_\tau}\left(\lambda_{\tau}^l+\mu\,\hat{\!\lambda}_{n}^l\mathrm{sign}(\,\hat{\!\lambda}_{\tau}^l)\right)
\end{bmatrix},&\quad\hat{\!\lambda}_{n}^l\leq0,\,|\,\hat{\!\lambda}_{\tau}^l|>-\mu\,\hat{\!\lambda}_{n}^l\,(\mbox{slip})\label{eq:mortar_stick_resi}\\
\delta W_{\mbox{\tiny nc}}^e = &
\begin{bmatrix}
 \delta\vec x_m^1\\[3pt]
 \delta\vec x_i^2\\[3pt]
 \delta\lambda_{n}^l\\[3pt]
 \delta\lambda_{\tau}^l\\
\end{bmatrix}^\intercal
\begin{bmatrix}
 0\\
 0\\
 -\frac{1}{\varepsilon_n}\lambda_{n}^l\\[10pt]
 -\frac{1}{\varepsilon_\tau}\lambda_{\tau}^l\\
\end{bmatrix},&\quad\hat{\!\lambda}_{n}^l>0\,(\mbox{no
contact}).\label{eq:mortar_nc_resi}
\end{align}
It is important to remark that the contact status is now based on the integral quantities (\ref{eq:integgn}, \ref{eq:gt_var}).
\begin{equation*}
 \hat\lambda_{n}^l = \lambda_{n}^l+\varepsilon_n\bar{g}_{n}^l,\quad
 \hat\lambda_{\tau}^l =
 \lambda_{\tau}^l+\varepsilon_\tau\mathring{\bar{g}}_{\tau}^l.
\end{equation*}

\section{Extended finite element methods}

\begin{figure}[htb!]
 \centering
 \includegraphics[width=1\textwidth]{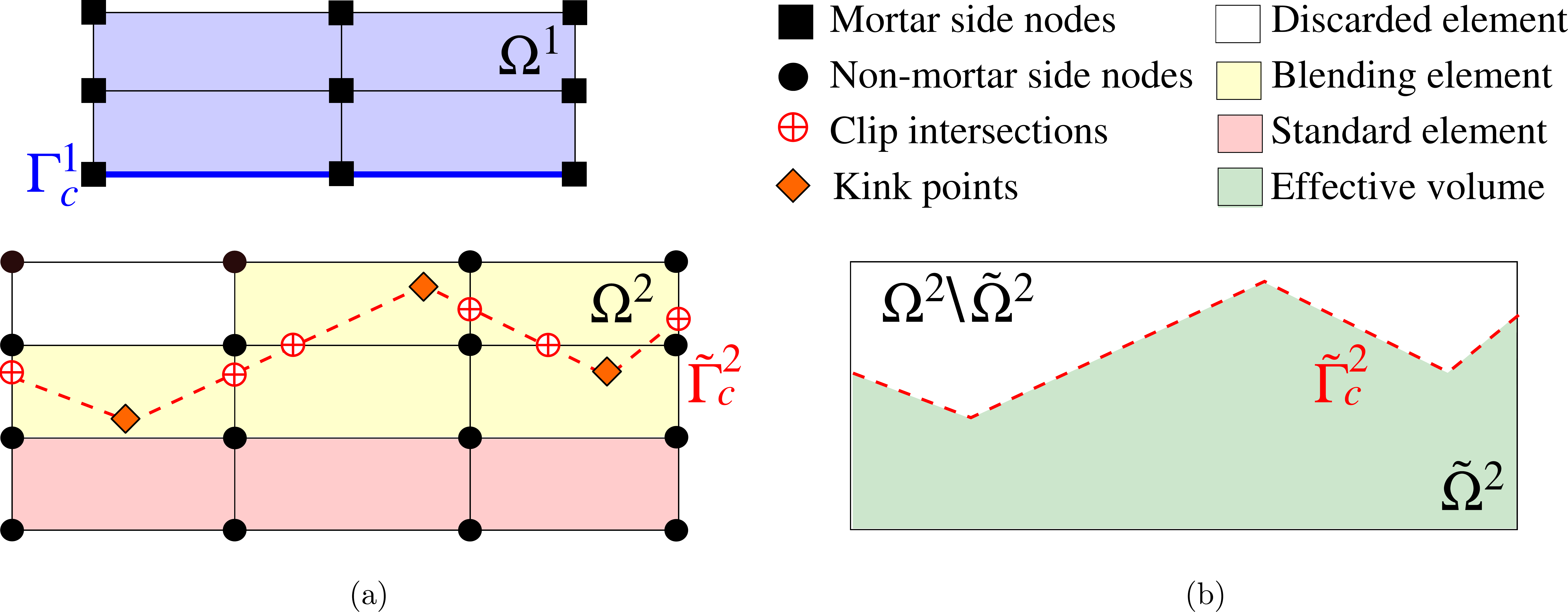}
 \caption{(a) Discrete contact interface pair: $\Gamma_{c}^1$ and $\Gamma_c^2$ denote real mortar surface and
 embedded non-mortar surface, respectively; (b) effective volume of the host domain $\tilde\Omega^2$.
 \label{fig:mortar_mortex_setup}}
\end{figure}

In this section, we present some elements of the extended finite element method (X-FEM)~\cite{dolbow_extended_1999,belytschko_review_2009}. In complement to the mortar method, X-FEM is a key ingredient required to construct a unified MorteX framework to treat contact problems along  along surfaces embedded in the bulk mesh. By analogy with the MorteX method for tying~\cite{akula_tying_paper}, the mesh hosting the embedded surface will be referred to as the host mesh. Such an embedded (or virtual) surface splits the domain of the host solid into two parts. One part is discarded (i.e. considered as empty space) and the remaining effective part represents the solid, see Fig.~\ref{fig:mortar_mortex_setup}. The virtual line $\tilde\Gamma_c^2$ runs through elements, which will be called blending elements. As such, these elements are also made of discarded and effective parts. The finite elements are then essentially partitioned into three distinct categories, namely the discarded, blending and standard elements. Here, X-FEM will be used  to account for the fact that only the effective volume $\tilde\Omega^2$ contributes to the internal virtual work.

The virtual surface $\tilde\Gamma^2_c$ of the host domain is treated as an internal
discontinuity, the outer part of the volume $\Omega^2\setminus\tilde\Omega^2$ is disregarded using the X-FEM framework.
The X-FEM relies on enhancement of the FEM shape functions used to
interpolate the displacement fields. The enrichment functions
describing the field behavior are incorporated locally into the finite element
approximation. This feature allows the resulting displacement to capture discontinuities.  
The subdivision of the host mesh is
defined by indicator function $\phi({\vec
X}): \mathbb R^{\mathrm{dim}}\to \{0,1\}$ (where ${\vec X}$ is
the spatial position vector in the reference configuration 
in domain $\Omega^2$)~\cite{sethian_level_1999}. The
indicator function is non-zero only in the efficient part of the host domain
$\tilde\Omega^2$: 
$$
  \phi(\vec X) = \begin{cases}
                    1,&\mbox{ if }\vec X\in \tilde{\Omega}^2;\\
                    0,&\mbox{ elsewhere.}
                 \end{cases}
$$
The discontinuity surface $\tilde\Gamma^2_c$ can be seen as a level-set defined as follows:
$$
  \tilde\Gamma^2_g = \left\{\vec X \in \Omega^2: \; \nabla\phi(\vec X) \ne 0\right\}.
$$
In the considered framework, for the sake of simplicity, we will assume that the embedded surface is represented by a piece-wise linear line with possible kink points occurring inside host elements.

\begin{figure}[htb!]
\centering
\includegraphics[width=\textwidth]{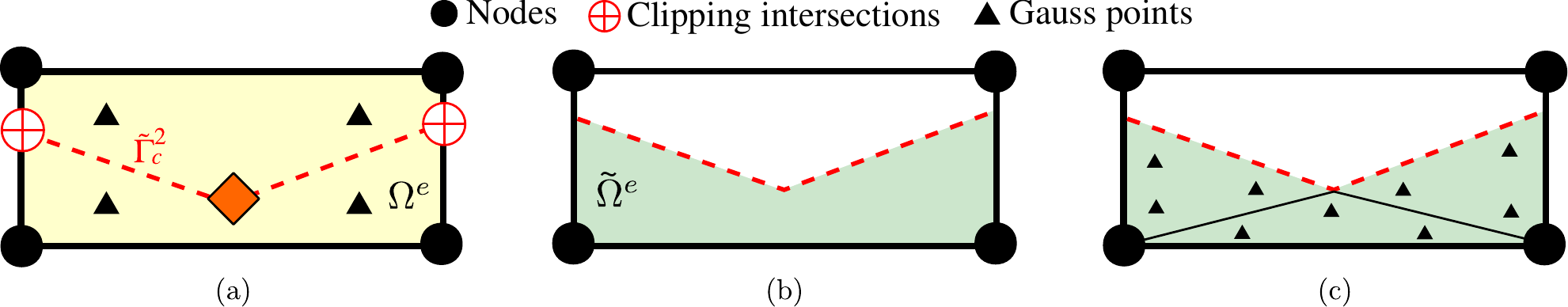}
\caption{Selective integration scheme for blending elements: (a) blending
element with standard Gauss points; (b) elemental volume segregation based on
polygon clipping (shaded part represents the effective volume); (c) triangulation of the effective volume, and
reinitialization of Gauss points, which is used to evaluate associated integrals using a Gauss quadrature rule.}
\label{fig:virtual_surface_partition}
\end{figure}

In practice, the enrichment of shape functions in case of void/inclusion problem
can be  simply replaced by a selective integration
scheme~\cite{sukumar_modeling_2001}.  For the standard elements, there is no
change in volume of integration and the discarded elements are simply excluded
from the volume integration procedure. In order to obtain the effective volume
of integration for each blending element, we
perform the clipping of the blending elements by the embedded surface
$\tilde\Gamma^2_c$ [Fig.~\ref{fig:virtual_surface_partition}] . The clipped element is virtually\footnote{Virtually stands here in the sense that the interpolation order and connectivity of nodes do not change.} triangulated to perform the integration of the internal virtual work using a Gauss quadrature rule.
In this framework, we use three newly initialized Gauss points per triangle which ensures a rather accurate integration in view of linear or bi-linear shape functions for triangular and quadrilateral elements, respectively.
The clipping of a single element by an arbitrary embedded surface results in one or several various polygons\footnote{Hereinafter, we assume that
all elements use first order interpolation, therefore all edges of elements are
straight. It enables us to assume that an intersection or difference of elements
can be always represented as one or several polygons.} either convex or
non-convex or even disjoint, which represent the effective volumes of integration. 
The tilde ($\tilde\bullet$) notations are adopted for the quantities related to
the effective part of the host mesh ($\tilde\Omega^2$).

\section{MorteX framework}\label{sec:mortex_framework}

Here, the MorteX framework, which is presented in a separate paper~\cite{akula_tying_paper} in the context of mesh tying problems, is extended to treat frictional contact problems. Similar to the tying
problem, it consists of two distinct procedures. The first procedure invokes the X-FEM method to account for the fact that only a portion of the body $\Omega^2$ is contributing to the internal virtual work; it is called the effective volume, see $\tilde\Omega^2$ on Fig.~\ref{fig:mortar_mortex_setup}. The second procedure deals with  the enforcement of the
contact constraints between a ``real'' surface $\Gamma_{c}^1$ (explicitly
represented by edges) and an embedded virtual surface $\Gamma^2_c$, which is geometrically
non-conformal with the finite element discretization and passing through the elements. The contact treatment is adjusted through few modifications of the
classical mortar contact schemes introduced in Section~\ref{sec:mortar_interface_disc}.

\subsection{MorteX interface discretization}\label{subsubsec:mortex_int_disc}

\begin{figure}[htb!]
   \centering
   \includegraphics[width=1\textwidth]{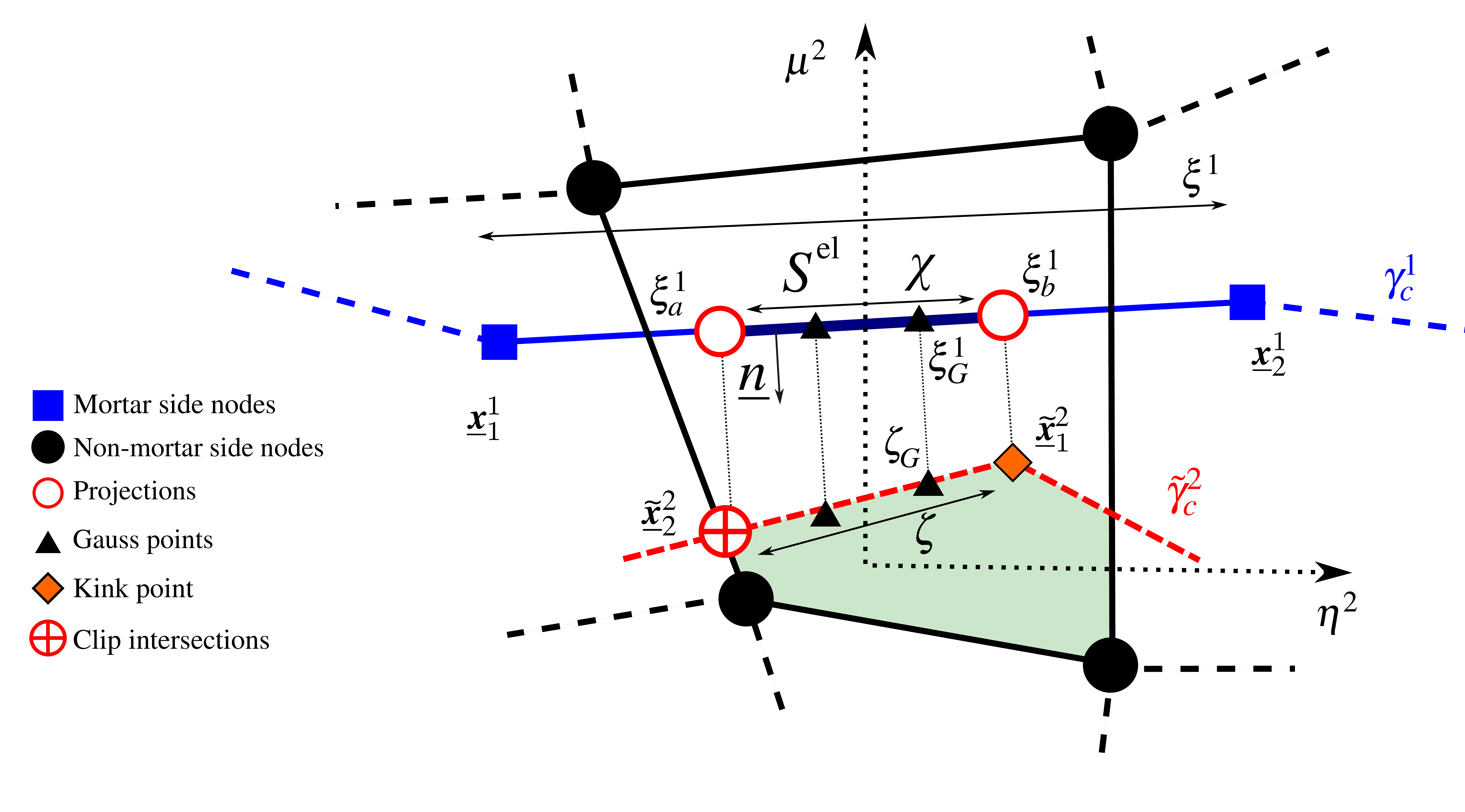}
   \caption{Example of mortar domain: the contact element is formed between a mortar side segment connecting $\vec x_1^1$ and $\vec x_2^1$ and a blending element; the integration is carried out over a portion of mortar segment and a corresponding portion of embedded surface (dashed line) between a clip intersection $\tilde{\vec{ x}}_2^1$ and a kink point $\tilde{\vec x}_1^2$.}
   \label{fig:mortx_ele}
\end{figure}

Similar to the tying problem in the MorteX framework, the rationale behind the choice of
mortar and non-mortar sides, holds for the contact problems as
well~\cite{akula_tying_paper}: the real surface is always selected as the mortar surface. 
The interpolations of  the nodal positions in the current configuration and the
Lagrange multipliers along the mortar side $\Gamma_c^1$ remain the same as for the classical
schemes, and are given by:
\begin{equation}
\vec x^1(\xi^1) = N_m(\xi^1)\vec x_m^1,\quad m\in[1,\mathrm{M}],
\label{eq:mortex_mort_interpol}
\end{equation}
\begin{equation}
    \lambda_{n,\tau}(\xi^1) = \Phi_l(\xi^1)\lambda^l_{n,\tau},\quad
    l\in[1,\mathrm{L}].
\label{eq:mortex_lambda_interpolations}
\end{equation}
where as previously $\mathrm M$ and $\mathrm L$ is the number of nodes per
mortar segment and those carrying Lagrange multipliers respectively. Vector $\vec
x_m^1$ represents the nodal positions of the mortar nodes forming the segment, and $\xi^1\in[-1;1]$ is the parametric coordinate of the mortar side. 
The parametrization on the non-mortar side requires more consideration. 
As shown in Fig.~\ref{fig:mortx_ele}, the embedded surface $\tilde\Gamma_c^2$ is divided
into straight segments, whose vertices can either be a clip intersection (i.e. the
intersection between $\tilde\Gamma_c^2$ and the edge of a bulk element of the
non-mortar side)  or a kink point (a vertex of the discrete virtual surface
$\tilde\Gamma_c^2$ that lies inside a blending element). Hereinafter, we will refer to these vertices as ``points of
interest'', without distinction between kink and clip intersection points; their notation will be equipped with a tilde $\tilde{\vec x}_j^2,  j \in [1,2]$. 
The shape of the embedded surface in the current configuration  parametrized by $\zeta \in [-1;1]$ between two points of interest, can be defined
using two-dimensional shape functions inherited from the host element [see Fig.~\ref{fig:mortx_ele}]:
\begin{equation}
{\vec x}^2(\zeta) =
N_i\left(\mu^2(\zeta),\eta^2(\zeta)\right)\vec x_i^2 
\label{eq:mortex_nmort_interpol}.
\end{equation}
Irrespective of the fact that the embedded surface is assumed piece-wise linear
in the reference configuration, it can become piece-wise non-linear in the
current configuration, therefore the displacement along every linear segment cannot be
parametrized by linear shape functions.
Note that here, in the MorteX contact set-up, the kink points represent vertices of the
discretized surface $\tilde\Gamma_c^2$, whereas they represented imprints of mortar side
nodes in case of MorteX tying set-up~\cite{akula_tying_paper}.

\subsubsection{MorteX contact element}
A MorteX contact element is formed between a single mortar segment and a blending element, and it consists of $(\mathrm M+\mathrm N)$  nodes, $\mathrm M$ from the
mortar segment, and $\mathrm N$ from blending element. In addition the contact
element stores $\mathrm L$ or $\mathrm L \times \mathrm{dim}$ Lagrange multipliers for the frictionless and
frictional cases, respectively.  
As was shown in~\cite{akula_tying_paper}, to avoid mesh-locking (or over-constraining
of the interface), which results in spurious oscillations, the number of Lagrange multipliers should be balanced by the number of
blending elements, i.e. often it is reasonable to use one set per blending
element, a detailed discussion can be found in Section~\ref{sec:stab_schemes}.

\subsubsection{Discrete integral kinematic quantities}\label{subsec:disc_integ_qtys_mortex}
The aforementioned selective integration scheme, used to accommodate for the presence of an embedded
virtual surface, leads to changes in the discrete contact integral quantities:
the integral gap vector~\eqref{eq:integral_gap_vector}, the integral normal
gap~\eqref{eq:integgn} and the incremental slip~\eqref{eq:gt_var}.   
For a MorteX contact element, these nodal
quantities are now evaluated along the interface formed by a pair
of real and embedded segments (tilde-notations are used):

\begin{align}
{\tilde{\bar{\! \vec g}}^{l}}&= \left[{D}_{lm}\vec x_m^1 -
\tilde{{M}}_{li}{\vec
x}_i^2\right],\label{eq:integral_gap_vector_mortex}\\\tag*{(integral gap
vector)}\\
\tilde{\bar g}_n^l &= \tilde{\bar{\vec g}}^l\cdot\vec
n,\label{eq:integral_gn_mortex}\\\tag*{(integral normal gap)}\\
    \mathring{\tilde{\bar{g}}}^l_\tau &=-\vec\tau\cdot
    \left[\left(D_{lm}(t_j)-D_{lm}(t_{j-1})\right){\vec
    x}_m^1(t_{j-1})-\left(\tilde{M}_{li}(t_j)-\tilde{M}_{li}(t_{j-1})\right){\vec
    x}_i^2(t_{j-1})\right]\label{eq:scal_integ_tang_gap_mortex}\\
\tag*{(nodal incremental slip)}
\end{align}
where $i\in[1,N]$, $l\in[1,L]$, $m\in[1,M]$, ${D}_{lm}$ and
$\tilde{M}_{li}$ are the mortar integrals in the
MorteX framework, which will be defined in~\eqref{eq:M_op_eval_mortex_contact}. Note that within the MorteX framework, the definitions of
purely mortar side quantities, like the
mortar segment normal $\vec n$ and mortar side integral $D_{lm}$ remain the same as in the classical mortar framework, whereas $\tilde M_{lm}$ now presents a convolution of volumetric and surface shape functions of primal and dual quantities, respectively.

\subsubsection{Discrete contact virtual work}
The MorteX residual vector  is larger than the standard mortar one as it involves the displacement DoFs from the bulk of the non-mortar side of the interface. The residuals for stick,
slip and non-contact statuses are:

\begin{align}
\delta \tilde W_{\mbox{\tiny stick}}^{\mathrm{el}} = &
\begin{bmatrix}
 \delta\vec x_m^1\\[5pt]
 \delta\vec x_i^2\\[5pt]
 \delta\lambda_{n}^l\\[5pt]
 \delta\lambda_{\tau}^l\\
\end{bmatrix}^\intercal
\begin{bmatrix}
\left(\hat{\!\lambda}_{n}^l\vec
 n+\hat{\!\lambda}_{\tau}^l\vec\tau\right)D_{lm}\\[5pt]
-\left(\hat{\!\lambda}_{n}^l\vec
 n+\hat{\!\lambda}_{\tau}^l\vec\tau\right)\tilde M_{li}\\[5pt]
 \tilde{\bar{g}}_{n}^l\\[5pt]
 \mathring{\tilde{\bar{g}}}_{\tau}^l\\
\end{bmatrix},&\quad\hat{\!\lambda}_{n}^l\leq0,\,|\,\hat{\!\lambda}_{\tau}^l|\leq-\mu\,\hat{\!\lambda}_{n}^l\,(\mbox{stick})\label{eq:mortex_stick_resi}\\
\delta \tilde W_{\mbox{\tiny slip}}^e = &
\begin{bmatrix}
 \delta\vec x_m^1\\[5pt]
 \delta\vec x_i^2\\[5pt]
 \delta\lambda_{n}^l\\[5pt]
 \delta\lambda_{\tau}^l\\
\end{bmatrix}^\intercal
\begin{bmatrix}
 \left(\hat{\!\lambda}_{n}^l\vec
n-\mu\,\hat{\!\lambda}_{n}^l
\mathrm{sign}(\,\hat{\!\lambda}_{\tau})\vec\tau\right)D_{lm}\\[5pt]
 -\left(\hat{\!\lambda}_{n}^l\vec
 n-\mu\,\hat{\!\lambda}_{n}^l\mathrm{sign}(\,\hat{\!\lambda}_{\tau})\vec\tau\right)\tilde
 M_{li}\\[5pt]
 \tilde{\bar{g}}_{n}^l\\[5pt]
-\frac{1}{\varepsilon_\tau}\left(\lambda_{\tau}^l+\mu\,\hat{\!\lambda}_{n}^l\mathrm{sign}(\,\hat{\!\lambda}_{\tau}^l)\right)
\end{bmatrix},&\quad\hat{\!\lambda}_{n}^l\leq0,\,|\,\hat{\!\lambda}_{\tau}^l|>-\mu\,\hat{\!\lambda}_{n}^l\,(\mbox{slip})\label{eq:mortex_stick_resi}\\
\delta \tilde W_{\mbox{\tiny nc}}^e = &
\begin{bmatrix}
 \delta\vec x_m^1\\[5pt]
 \delta\vec x_i^2\\[5pt]
 \delta\lambda_{n}^l\\[5pt]
 \delta\lambda_{\tau}^l\\
\end{bmatrix}^\intercal
\begin{bmatrix}
 0\\
 0\\
 -\frac{1}{\varepsilon_n}\lambda_{n}^l\\[5pt]
 -\frac{1}{\varepsilon_\tau}\lambda_{\tau}^l\\
\end{bmatrix},&\quad\hat{\!\lambda}_{n}^l>0\,(\mbox{no
contact}).\label{eq:mortex_nc_resi}
\end{align}

In addition, the contact detection as well as the evaluation of the
residual vector are based on the following modified weighted
integral quantities~\eqref{eq:integral_gap_vector_mortex}, 
\eqref{eq:integral_gn_mortex} and \eqref{eq:scal_integ_tang_gap_mortex}.
\begin{equation*}
 \hat\lambda_{n}^l = \lambda_{n}^l+\varepsilon_n\tilde{\bar{g}}_{n}^l,\quad
 \hat\lambda_{\tau}^l =
 \lambda_{\tau}^l+\varepsilon_\tau\mathring{\tilde{\bar{g}}}_{\tau}^l.
\end{equation*}

\subsubsection{MorteX integral evaluation}
The numerical procedures of the classical mortar framework to evaluate the
mortar integrals need to be adapted for the MorteX framework similar to what was
done for the tying problem. 
However, the numerical procedure
adapted for the tying problem in MorteX framework cannot be used directly for the
contact problems, because the former does not require any
projections between the tying surfaces $\tilde{\Gamma}_g^2$ and
$\Gamma_g^1$~\cite{akula_tying_paper}.
The lack of projections in the tying framework is due to the fact that the embedded surface is nothing but an imprint of the mortar surface of the patch domain
 in the reference configuration.  However, for
the non-linear contact problem, where the contact interface
continuously changes in response to the deformations the bodies, the conformity of surface meshes cannot be ensured.  It implies a need
for a projection step similar to the classical mortar framework to determine the
    limits of integration mortar-domain $S^{\mathrm{el}}$ (see Fig.~\ref{fig:mortx_ele}).  The quantities
    projected are points of interest rather than non-mortar segment nodes,
    which is the case in the classical mortar.  The mortar projection coordinates for the extremities of the integration ($\xi^1_a,\xi^2_b$) are found
    by solving the following equation:
\begin{equation} 
    { \left[N_m(\xi^1)\vec x^1_m - \tilde{\vec x}^2_i\right]\times \vec n = 0}
    \label{eq:mortex_xi_proj},
\end{equation}
where $\tilde{\vec x}^2$ denote points of interest (kink or clip intersection). The resulting
projections $\xi_{a/b}^1$ serve as limits of the mortar domain $S^{\mathrm{el}}$,  which is parametrized
by $\chi\,\in[-1,1]$  (see Fig.~\ref{fig:mortx_ele}). 
To evaluate the integrals using Gauss quadrature, the mortar-side Gauss points
$\xi^1_G$ are projected along the mortar-segment normal $\vec n$
onto the non-mortar side. The corresponding local coordinates $\zeta_G$ are determined by solving for each Gauss point $\xi^1_G$:
\begin{equation}
    \left[N_i^2(\mu^2(\zeta_G),\eta^2(\zeta_G)){\vec x}^2_i - N_m(\xi^1_G)\vec x^1_m\right]\times \vec n = 0.
\end{equation}
The Gauss point location obtained in the parametric space $\zeta$ is mapped onto
the underlying parameterization of the non-mortar element ($\mu^2,\eta^2$).  
\begin{equation}
    \mu_{{G}}^{2}{(\zeta_G)} =
    \frac{1}{2}(1-\zeta_G)\mu_{1}^2\,+\,\frac{1}{2}(1+\zeta_G)\mu_{2}^2,\quad
    \eta_{{G}}^{2}{(\zeta_G)} =
    \frac{1}{2}(1-\zeta_G)\eta_{1}^2\,+\,\frac{1}{2}(1+\zeta_G)\eta_{2}^2,
\end{equation}
where $(\mu_{j}^2,\eta_{j}^2)$ are the coordinates of the points of interest
$\tilde{\vec x}^2_j \; j\in[1,2]$ in the parent domain.
The integrals are evaluated as:
\begin{align}
    D_{lm} &=
    \int\limits_{S^{\mathrm{el}}}\Phi_l(\xi^1)N_m^1(\xi^1)\,d\Gamma=
    \sum_{G=1}^{N_G} w_{G}\Phi_{l}(\xi^{1}_{G})N_{m}^{1}(\xi^{1}_{G})J_{\mbox{\tiny seg}}(\xi_{G}^{1}) 
    \label{eq:D_op_eval_mortex_contact}\\
    \tilde{{M}}_{li}&=\int\limits_{S^{\mathrm{el}}}\Phi_l(\xi^1)N_i^2(\mu^2(\zeta),\eta^2(\zeta))\,d\Gamma
    =\sum_{G=1}^{N_G}  w_{G}\Phi_{l}(\xi^{1}_{G})N_{i}^{2}\big(\mu^{2}_{G},\eta^{2}_{G}\big)J_{\mbox{\tiny
    seg}}(\xi_{G}^{1})
    \label{eq:M_op_eval_mortex_contact}
\end{align}
where the Jacobian is given by:
\begin{equation}
    J_{\mbox{\tiny seg}}(\xi_{G}^{1}) =
    \bigg|\frac{\partial{N^1_i}}{\partial\xi^{1}} \frac{\partial \xi^1}{\partial \chi}\vec
    x_{i} \bigg|.
    \label{eq:2D_jacob_mtx}
\end{equation}

\section{Stabilization of the MorteX method\label{sec:stab_schemes}}

As known, mixed formulations for interface problems are frequently prone to mesh locking phenomenon causing spurious oscillations near the interface~\cite{sanders_nitsche_2012}. 
In the particular case of contact and mesh tying, the mesh locking can manifest itself if the number of constraints surpasses the number of available degrees of freedom to fulfil them.
Therefore, interfaces with high contrast of mesh densities across the interface are more likely to be prone to mesh locking.
Especially, this manifestation is more severe for high elastic (or stiffness) contrasts between contacting or tied domains~\cite{akula_tying_paper}. For example, prescribing Dirichlet boundary condition
along an embedded surface, which is known to pose similar problems in the X-FEM framework~\cite{barbosa_finite_1991,bechet_stable_2009,ramos_new_2015}, can be seen as a particular case of infinite material contrast across the interface. 
In order to avoid spurious oscillations at contact interfaces between materials
of high contrast or meshes of too different densities, a stabilization technique should be used. 
In~\cite{akula_tying_paper} we suggested to coarse-grain Lagrange multipliers in order to avoid over-constraining of the interface.
This technique was proved particularly efficient and easy to use. In contrast to Nitsche method which needs to be stabilized in similar situations accordingly to the material contrast~\cite{sanders_nitsche_2012}, a simple mesh-based spacing parameter is used in the coarse-graining of Lagrange multipliers. Therefore, an \emph{automatic} choice of the coarse-graining parameter (either local or global one) is straightforward to integrate in a finite element procedure.
Here, the coarse graining technique will be briefly recalled and tested on a simple contact patch test similar to~\cite{taylor1991patch,el_abbasi_stability_2001}.
An alternative approach to address the mesh locking is the triangulation of blending elements, which was also studied in~\cite{akula_tying_paper}. Even though the triangulation does not resolve the problem of over-constraining, it ensures a linearity of the displacement field along any straight line inside blending elements and can potentially help for some problems. Nevertheless, as was shown in~\cite{akula_tying_paper}, this triangulation allows to get rid from the spurious oscillations only in a very limited number of cases, among which the classical contact patch test~\cite{taylor1991patch,el_abbasi_stability_2001}.

\subsection{Coarse-grained interpolation of Lagrange multipliers}\label{sec:cgi}

\begin{figure}[htb!]
\centering
\includegraphics[width=\textwidth]{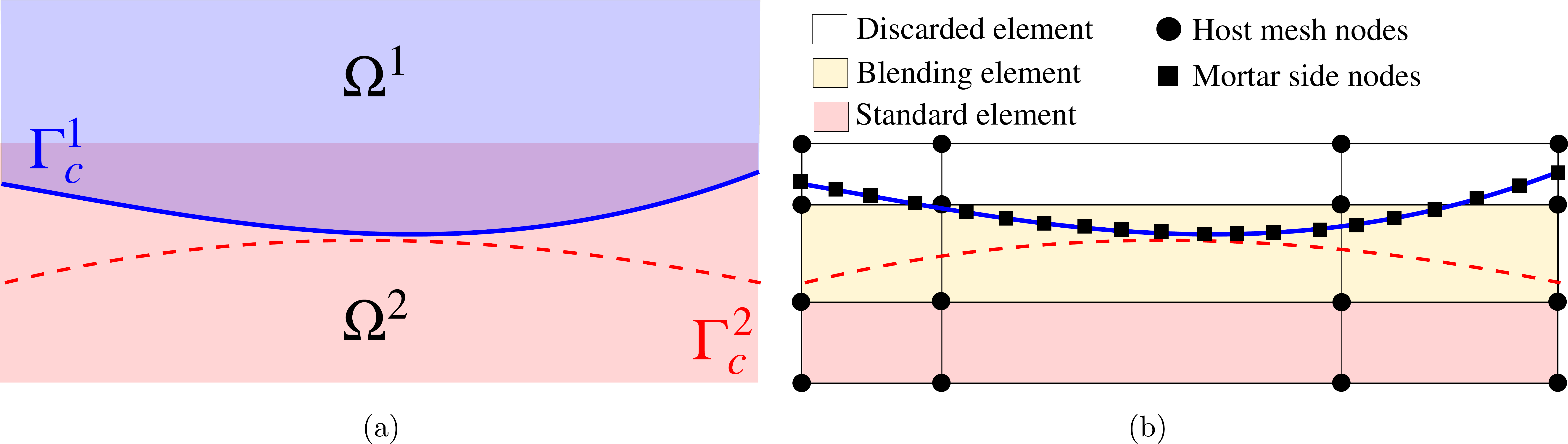}
\caption{Contact between real surface $\Gamma_c^1$ and embedded surface
  $\Gamma^2_c$: (a) continuum set-up, (b) discretized framework (only surface discretization is shown on the mortar side, whereas the inherent discretization of the embedded surface is not shown on purpose to avoid any confusions).}
\label{fig:stabilization_prob_setting}
\end{figure}

\begin{figure}[htb!]
\centering
\includegraphics[width=1\textwidth]{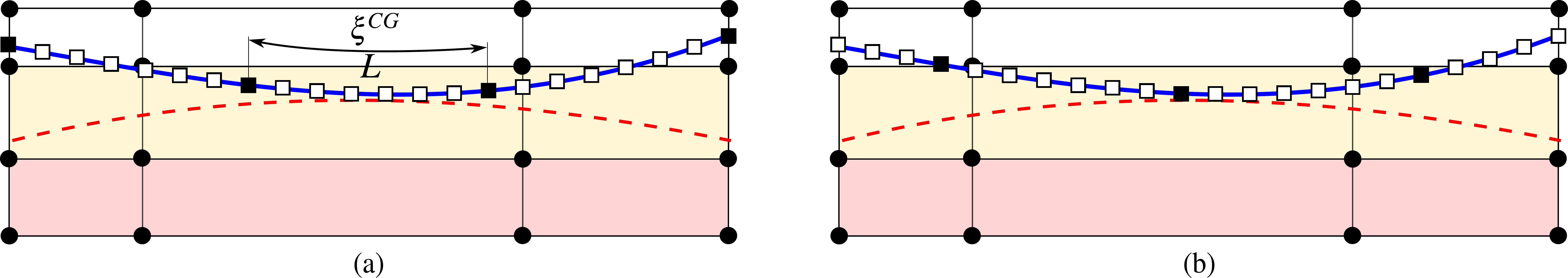}
\caption{Coarse graining of Lagrange multipliers for an interface with mesh contrast parameter $m_c=7$ (the ratio of 21 mortar segments to 3 blending elements): (a) a global approach in which every seventh mortar node is selected as master one and is assigned to carry Lagrange multipliers, (b) local coarse-graining in which  only one mortar node per homologue blending element is selected as master.}
\label{fig:coarse_grain_concept}
\end{figure}

The concept of coarsening, inspired from~\cite{moes_imposing_2006,bechet_stable_2009} and implemented in~\cite{akula_tying_paper}, consists in selecting only few (master) nodes of the mortar surface to carry Lagrange multipliers. Therefore the associated traction field is interpolated over the mortar surface using classical shape functions, with the difference that they span now few elements. Alternatively, one can assign Lagrange multipliers to remaining (slave) mortar nodes and use classical local interpolation shape functions for Lagrange multipliers; however, the values of these slave Lagrange multipliers should be constrained to be a linear combination of Lagrange multipliers of neighbouring master nodes, the associated weights still should be defined by non-local shape functions spanning all the edges which connect the corresponding master nodes. This alternative approach, which is used in this work, can be achieved using multi-point constraints (MPC) for Lagrange DoFs. The concept of coarse-grained interpolation (CGI) is illustrated in Fig.~\ref{fig:coarse_grain_concept}.
Here, we choose an arbitrary discretized curved mortar surface $\Gamma^1_c$ which comes in contact with an embedded  surface $\Gamma^2_c$ cutting through
 blending elements of the non-mortar host mesh $\Omega^2$. In this example, on purpose we choose such discretization that mortar nodes (potentially carrying Lagrange multipliers) largely outnumber blending elements.
 In this situation, the related contact (or tying) interface can be considered over-constrained especially when the mortar side material is considerably more rigid than this of the non-mortar side.
 
To construct a coarse-grained interpolation, we proceed to the selection of master nodes on the mortar side, which will carry lagrange multipliers. This selection can be guided either by the global or a local mesh density contrast.
In the former case, we introduce a global mesh density contrast parameter $m_c$ which is the ratio of mortar-side segments $N^m$ to the number of blending elements $N^b$, i.e. $m_c = N^m/N^b$. 
For approximately equal discretizations $m_c \approx 1$ or coarser discretization of the mortar $m_c<1$, the coarse-graining is usually not needed. But for higher density on the mortar side $m_c > 1$,  only few mortar nodes should carry Lagrange multipliers. To characterize it, we introduce a coarse-graining spacing parameter or simply spacing parameter $\kappa$ , which denotes that only every $\kappa$-th mortar node will be assigned as master node [Fig.~\ref{fig:coarse_grain_concept}(b)]. The limit case $\kappa=1$, all mortar nodes carry Lagrange multipliers, which corresponds to the standard Lagrange interpolation (SLI). 
Evidently, such an approach based on the global mesh-density contrast, is expected to perform well only for regularly meshed surfaces and regular size of blending elements.
Otherwise, a local approach should be used, in which, for example, one master node per homologue blending element is selected [Fig.~\ref{fig:coarse_grain_concept}(b)].
The global approach results in a uniform spacing between master nodes (possibly with an exception of one extremity of the mortar side) and local approach has a potentially variable spacing.

Let assume that we have a pair of neighbouring master nodes $1$ and $N$, carrying Lagrange multipliers $\vec\lambda_{1}$ and $\vec\lambda_N$ and separated by distance $L$, which is the sum of the lengths of segments $L_{i,i+1}$ with $i\in[1,N-1]$ connecting these nodes, i.e. $L = \sum_{i=1}^N L_{i,i+1}$. Therefore, within the context of MPC, to interpolate Lagrange multipliers we assign to every slave node $i$ a nominal value of Lagrange multiplier as
$$\tilde{\!\!\,\lambda}_i = \lambda_1 (1- L_{1,i})/L + \lambda_N L_{i,N}/L,$$ 
where $L_{i,j} = \sum_{k=i}^{j-1} L_{i,i+1}$ is the total length of segments connecting nodes $i$ and $j$. Subsequently, the classical local shape functions can be used to interpolate the field of Lagrange multipliers over every segment; this approach is adopted here.
Alternatively, a coarse-grained parametrization with normalized parameter $\xi^{CG}$ could be used; it should fulfil the following condition 
$$
  \forall i \in [1,N-1]:\quad \frac{\xi^{CG}_{i+1} - \xi^{CG}_{i}}{L_{i,i+1}} = \mathrm{const},
$$
then  classical but non-local shape functions can be used to interpolate the field of Lagrange multipliers:
\begin{equation}
  \lambda(\xi^{CG}) = \lambda^1 \tilde\Phi_1(\xi^{CG}) + \lambda^N \tilde\Phi_2(\xi^{CG}).
  \label{eq:cgi_lam_interpol}
\end{equation}

The choice of optimal spacing parameter $\kappa$ was studied in~\cite{akula_tying_paper}. To summarize the drawn conclusions: (1) the choice of the spacing, which ensures that the distance between master mortar nodes is greater than the average dimension of blending elements, allows to get rid of spurious oscillations; (2) to preserve the accuracy of the solution the $\kappa$ parameter should not be too large.

\subsection{Contact patch test}

\begin{figure}[htb!]
\centering
\includegraphics[width=\textwidth]{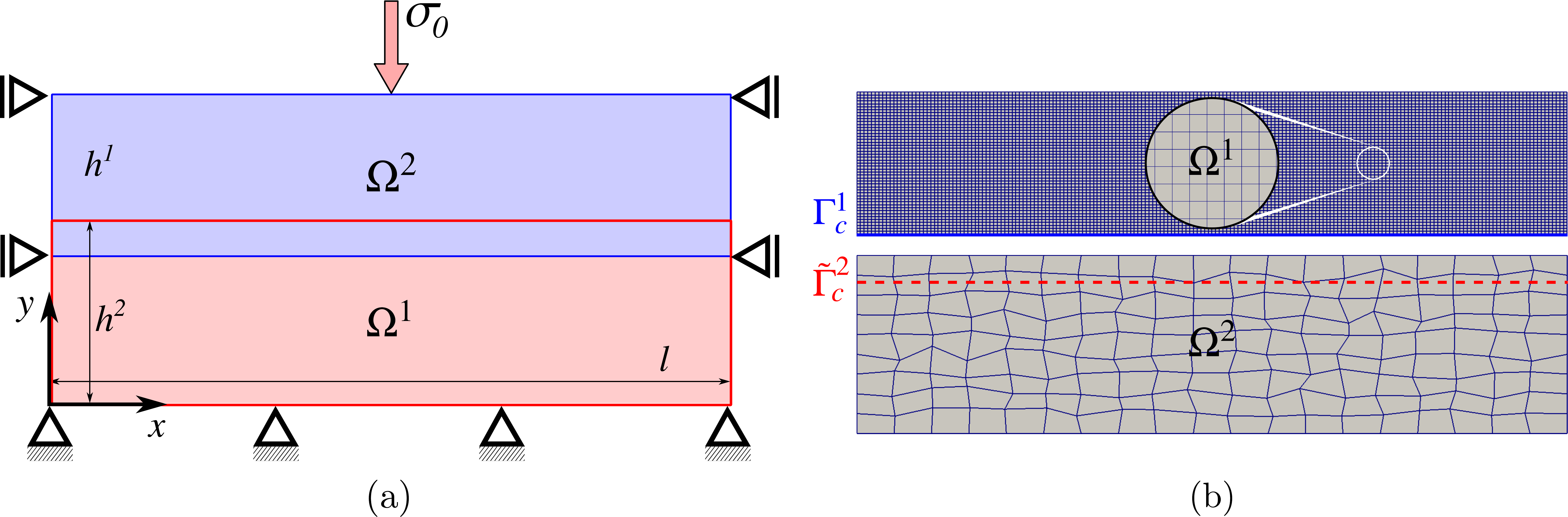}
  \caption{(a) Patch test setup; (b) host mesh has a coarser discretization than the mortar side mesh with the mesh density contrast $m_c\approx12$.}
\label{fig:patch_test_setup}
\end{figure}

To illustrate the stabilization techniques, a contact patch test shown in Fig.~\ref{fig:stabilization_prob_setting} is used, where outer surface $\Gamma_c^1 \subset \partial\Omega^1$ comes in contact with an embedded  surface
 $\Gamma_c^2 \subset \Omega^2$ (Fig.~\ref{fig:patch_test_setup}). The classical contact patch test~\cite{taylor1991patch,el_abbasi_stability_2001} is adjusted to the case of different materials by constraining the normal displacement on lateral sides. A very similar test was presented for mesh tying along straight interfaces in~\cite{akula_tying_paper}, with the only difference that here the developed MorteX \emph{contact} algorithm is used.

\begin{figure}[htb!]
\centering
\includegraphics[width=\textwidth]{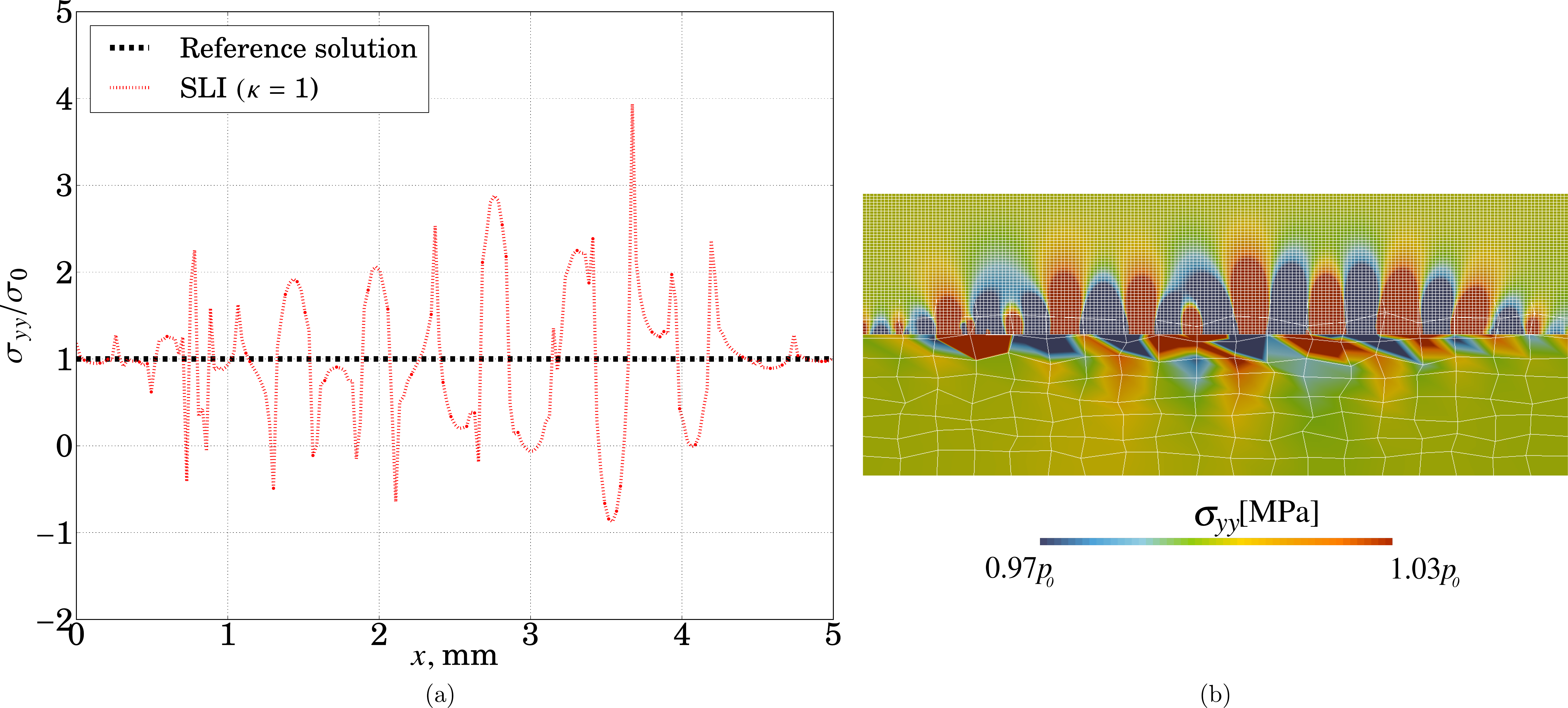}
\caption{Result of the patch-test in the absence of any stabilisation: (a) $\sigma_{yy}$ along $\Gamma_c^1$, (b) contour plot of $\sigma_{yy}$.}
\label{fig:patch_test_BAD_sol}
\end{figure}

The particular case of softer and coarser host mesh is chosen, as this combination is known to be the most prone to spurious oscillations along
the interface~\cite{akula_tying_paper}. Additionally, the host domain is meshed on purpose with
distorted quadrilateral elements in order to obtain a local parametrization that is not aligned with the embedded surface.
 Linear elastic material
properties are used for both the domains $\Omega^1$ ($E_1 = 1000 $ GPa, $\nu_1=0.3$) and $\Omega^2$
($E_2 = 1$ GPa, $\nu_2 = 0.3$).
The geometric set-up is illustrated in Fig.~\ref{fig:patch_test_setup}(a). The discretization for the domains is shown
in Fig.~\ref{fig:patch_test_setup}(b), resulting in a mesh-density contrast $m_c\approx 11$. Plane strain formulation is used.
A uniform pressure $\sigma_0=1$ Pa is applied on the top surface, while the bottom surface is
fixed in all directions ($\vec u=0$) and the lateral sides are constrained in the normal
direction $u_x=0$ (Fig.~\ref{fig:patch_test_setup} (a)).  The exact solution
for the vertical stress component is a uniform field $\sigma_{yy} = \sigma_0$. Fig.~\ref{fig:patch_test_BAD_sol} shows that the  standard Lagrange multiplier interpolation (SLI) exhibits high-amplitude spurious oscillations with errors locally exceeding $300$ \% and spanning many layers of elements in the bulk. 

\begin{figure}[htb!]
\centering
\includegraphics[width=\textwidth]{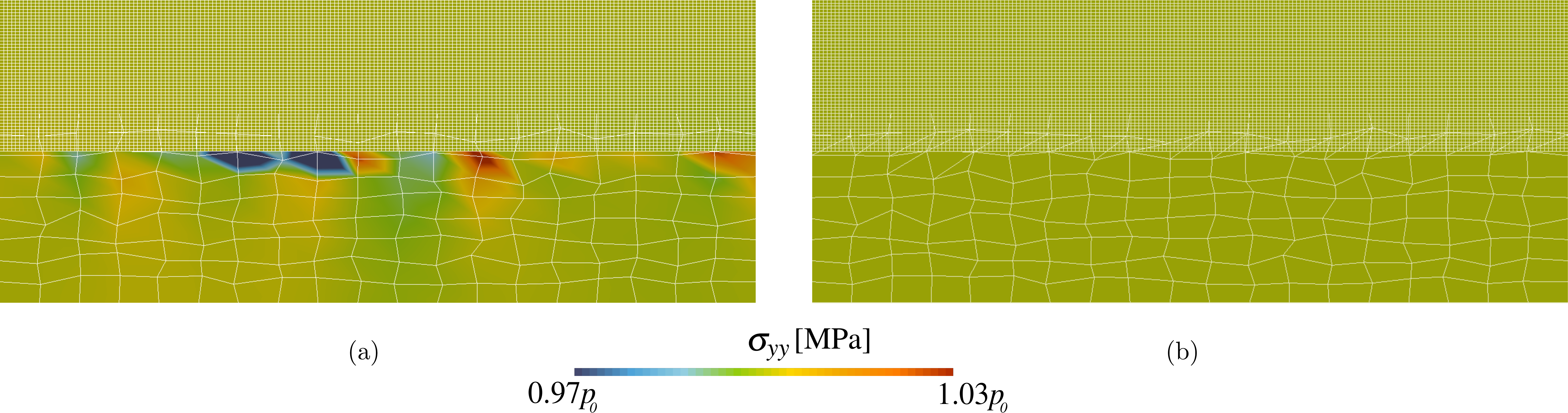}
\caption{MorteX patch test, stress component $\sigma_{yy}$: (a) coarse-grained interpolation (CGI) with two master nodes assigned on the extremities of the mortar surface;
(b) triangulation of blending elements.}
\label{fig:patch_test_stresses}
\end{figure}

As we seek to approximate a uniform stress state, a single Lagrange multiplier with a zero order interpolation should be sufficient. Here, however, two Lagrange multipliers are used to demonstrate the effect of coarse-graining interpolation. This interpolation is achieved by assigning Lagrange multipliers to both end nodes of
the mortar surface $\Gamma_c^1$ which act as master nodes for the multi-point
constraints, while the nodes lying in-between are the slaves, and the
Lagrange multipliers are expressed as a linear combination  of the Lagrange
multipliers hosted by the master nodes~\eqref{eq:cgi_lam_interpol}. 
Fig.~\ref{fig:patch_test_stresses}(a) shows that
the amplitude of the stress oscillations 
along the contact interface are significantly reduced ($< 1$  \%). 
However, the stresses $\sigma_{yy}$ in the host mesh still exhibit oscillations of
amplitude ($\approx 3$ \%). This shows that although the CGI scheme allows drastic reduction in stress oscillations, it is not sufficient to recover the
reference uniform stress distribution.
Surprisingly, the reference field can be recovered if initially quadrilateral blending elements of the host mesh are triangulated. 
As seen in Fig.~\ref{fig:patch_test_stresses}(b) no visible oscillations are observed in the interface even though this triangulation does not resolve the problem of over-constraining.

Nevertheless, as will be demonstrated below, passing this patch test does not guarantee, in the general case, that using triangular host mesh ensures accurate stress fields near the interface.
As was demonstrated in~\cite{akula_tying_paper} for mesh tying, the considered configuration of the patch test is the only example in which triangulation is sufficient to get rid of spurious oscillations and recover the reference solution.
In a more general case, when the interface stress is non-uniform, simple triangulation does allow to obtain accurate solution and is as prone to spurious oscillations as a quadrilateral mesh. 
Therefore, it could be reasonably thought that the coarse graining should be always combined with the triangulation of blending elements in order to ensure a good performance in any case. However, as was demonstrated in numerous examples in~\cite{akula_tying_paper}, the triangulation can slightly deteriorate the accuracy of the solution. Finally, from the practical point of view, the triangulation of blending elements can be a user-defined option, whereas coarse-grained interpolation is a needed feature as will be shown below on more realistic contact examples.

\section{Numerical examples}\label{sec:examples}
In this section, we carefully selected numerical tests to demonstrate the
accuracy, the robustness and the efficiency of the method. Linear elements are
used in all the examples. All the examples are set-up using both boundary-fitted
domains and embedded surfaces, solved within classical mortar and MorteX
frameworks, respectively. Both methods were implemented  in
in-house finite element suite Z-set~\cite{besson_large_1997}.
In the figures displaying contour stress plots,
transparency is used to visually differentiate between the ``discarded'' (most
transparent), ``blending'' (semi transparent) and ``standard'' (opaque)
elements. 

\subsection{Frictionless contact of cylinders}\label{subsec:hertz_sol}
\begin{figure}[htb!]
   \centering
   \includegraphics[width=\textwidth]{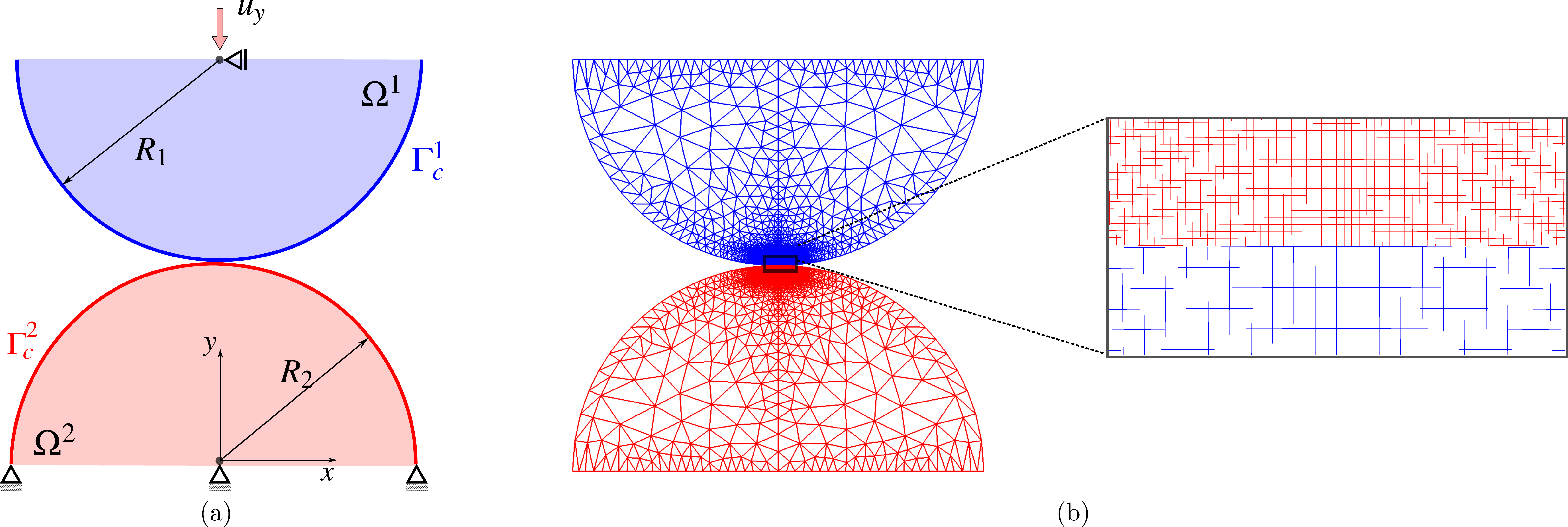}
   \caption{Mortar Hertzian contact: (a) problem set-up and boundary conditions; (b)
   FE discretization, with mesh contrast parameter $m_c\approx3$
   (zoom at the interface mesh).}
   \label{fig:hertz_fricless_setup_mortar}
\end{figure}

\begin{figure}[htb!]
   \centering
   \includegraphics[width=\textwidth]{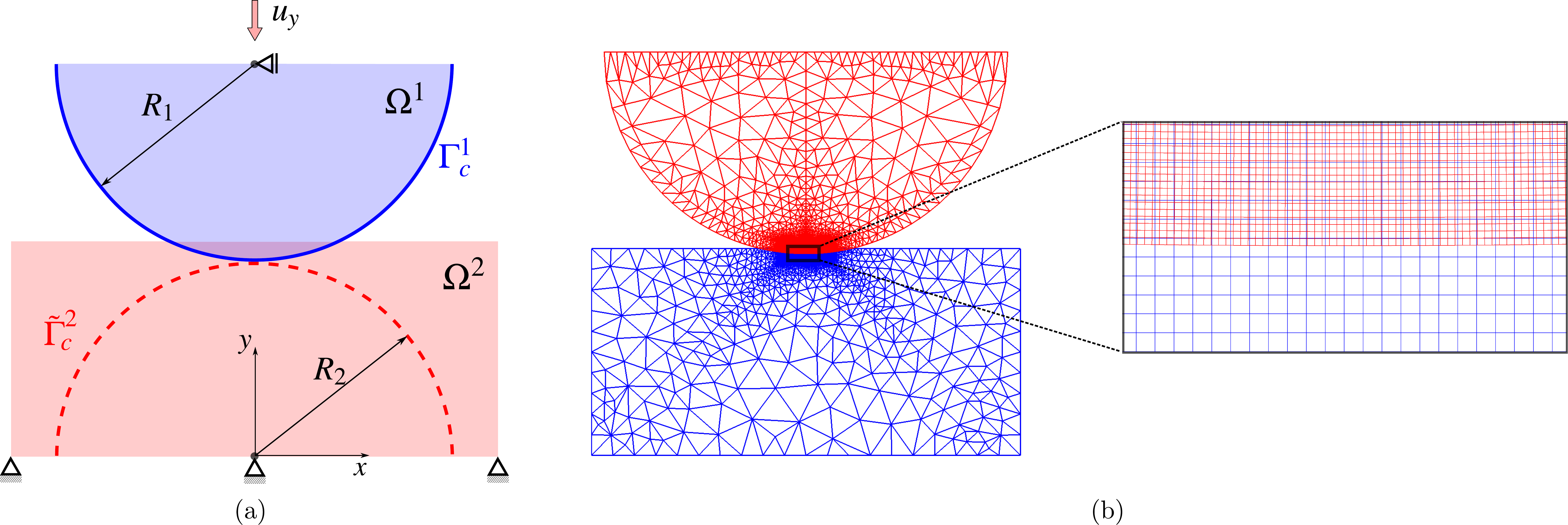}
   \caption{MorteX Hertzian contact: (a) problem set-up and boundary conditions; (b)
   FE discretization, with mesh contrast parameter $m_c\approx3$
   (zoom at the interface mesh).}
   \label{fig:hertz_fricless_setup}
\end{figure}

As demonstrated in~\cite{akula_tying_paper} for mesh tying problems, 
set-up involving a host that is softer and coarser than
the patch tend to exhibit spurious interface oscillations as a result of
mesh-locking. In order to illustrate the mesh-locking effect within the 
context of contact problems, we choose the below set-ups for mortar
[Fig.~\ref{fig:hertz_fricless_setup_mortar}] and MorteX
[Fig.~\ref{fig:hertz_fricless_setup}].
The problem under consideration is the frictionless Hertzian contact between two
infinite cylinders. For the MorteX set-up, the bottom
cylinder surface is embedded into a host rectangular domain $\Omega^2$. The
cylinders are of equal radii  $R_1=R_2=8$ mm. Linear
elastic materials are used for both domains $\Omega^1$ ($E_1,\nu_1$) and
$\Omega^2$ ($E_2,\nu_2$).
A material contrast is introduced by choosing $E_1/E_2=100$. 
The same Poisson
ratio of $\nu_1=\nu_2=0.3$ is used for both the domains. 
The top cylinder has a finer discretization compared to the bottom cylinder,
with the mesh contrast parameter $m_c\approx3$. 
A vertical displacement $u_y=0.005$ mm is applied on the surface of the top
cylinder. This results in a total reaction force of $P\approx0.016512$ N. 
The bottom surface is fully fixed. The middle
point of the top surface is fixed in $x$ direction. For the classical mortar
framework contact is enforced between the real-real surface pair
($\Gamma_c^1/\Gamma_c^2$), while for MorteX the contact is enforced
between the real-virtual surface pair ($\Gamma_c^1/\tilde\Gamma_c^2$).

The analytical solution for this problem is
derived from the Hertzian contact formulae for two cylinders, which defines the
maximum contact pressure ($p_{0}$), the semi-width of contact zone $a$, and the contact
pressure distribution $p$ along the $x$ coordinate~\cite{johnson_contact_1985}:

\begin{tabularx}{\textwidth}{XXX}
\begin{equation}
    p_0 = \sqrt{\frac{PE^*}{\pi R^*}},
    \label{eq:hertz_p_max}
\end{equation}&
\begin{equation}
    a = \sqrt{\frac{4PR^*}{\pi E^*}},
    \label{eq:hertz_con_width}
\end{equation}&
\begin{equation}
    p = p_0\sqrt{1-\bigg(\frac{x}{a}\bigg)^2}.
    \label{eq:hertz_distribution}
\end{equation}
\end{tabularx}
In the above equations, the effective elastic modulus $E^*$ is defined as
\begin{equation}
    E^* = \frac{E_1E_2}{E_1(1-\nu_2^2)+E_2(1-\nu_1^2)}
    \label{eq:hertz_E_effective}
\end{equation}
and the effective radius $R^*$, is evaluated as:
\begin{equation}
    R^*=\frac{R_1R_2}{R_1+R_2}.
    \label{eq:hertz_R_effective}
\end{equation}

\begin{figure}[htb!]
   \centering
   \includegraphics[width=\textwidth]{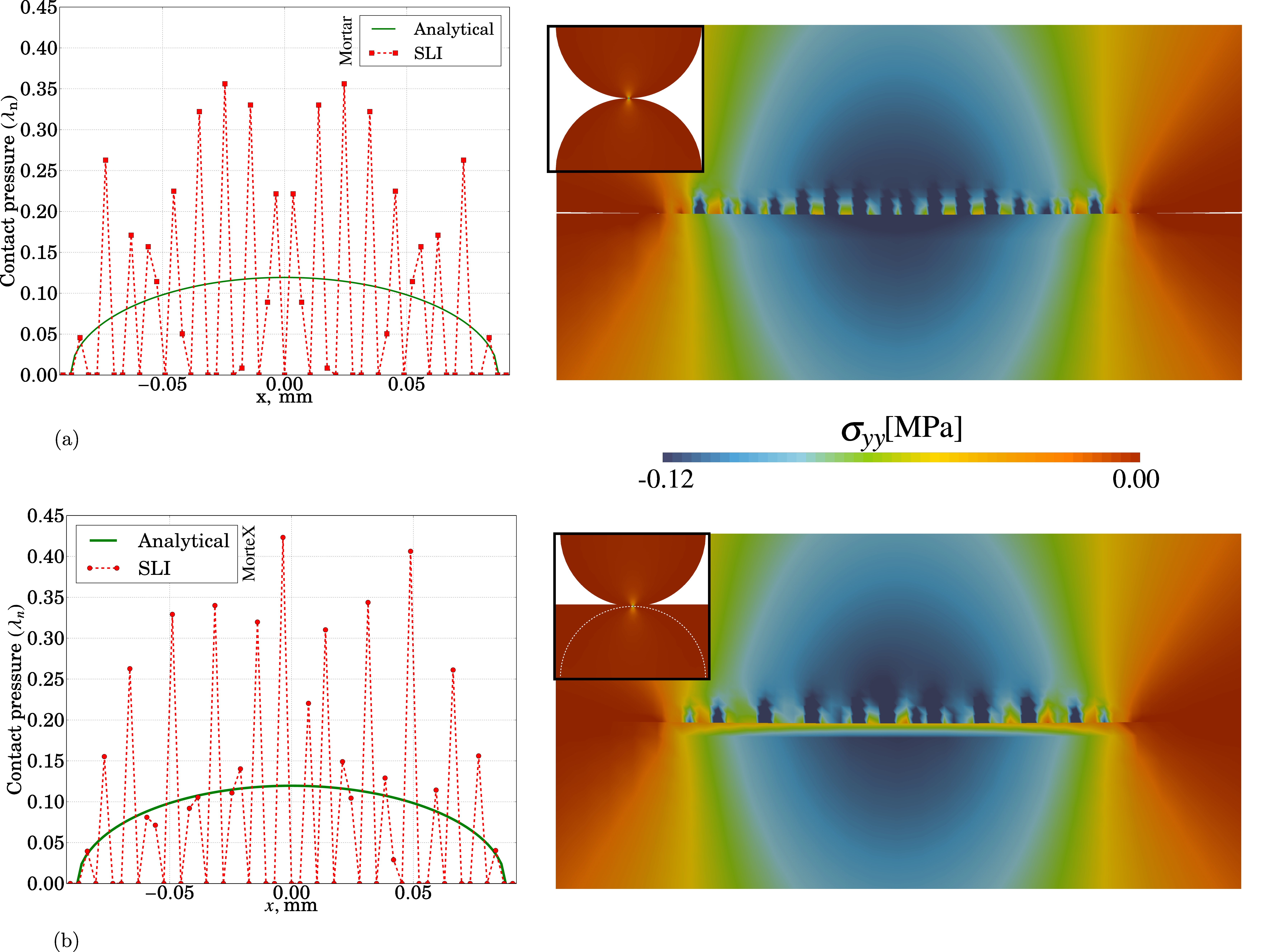}
   \caption{Contact stresses ($\lambda_n, \sigma_{yy}$) for the standard Lagrange multiplier interpolations
   (SLI): (a) Mortar; (b) MorteX methods.}
   \label{fig:hertz_fless_sli_results}
\end{figure}

\begin{figure}[htb!]
   \centering
   \includegraphics[width=\textwidth]{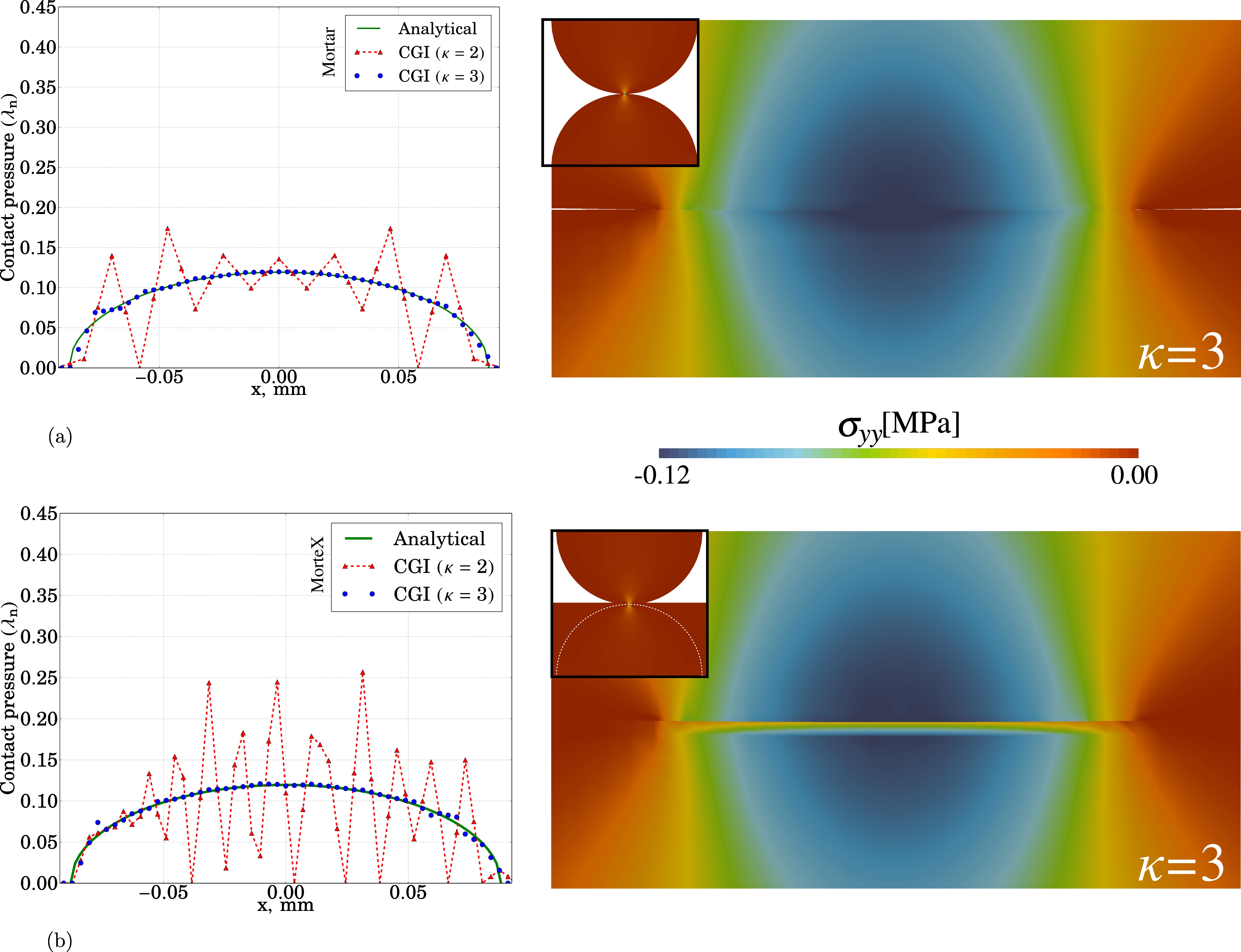}
   \caption{Contact stresses ($\lambda_n, \sigma_{yy}$) for the coarse grained Lagrange multiplier interpolations
   (CGI): (a) Mortar; (b) MorteX methods.}
   \label{fig:hertz_fless_cgi_results}
\end{figure}

Using the standard Lagrange multiplier interpolation in which every mortar side
node holds a Lagrange multiplier results in spurious
oscillations in the stresses [see Fig.~\ref{fig:hertz_fless_sli_results}]. Similar oscillations have been demonstrated in~\cite{akula_tying_paper} in mesh tying problems. 
The stabilization technique, coarse graining of
Lagrange multiplier interpolations [see Section~\ref{sec:cgi}], was proved
efficient in removing these oscillations on
various problem settings for mesh tying problems. We adapt the same
stabilization approach for contact problems, to both the mortar and MorteX frameworks.
The plots in Fig.~\ref{fig:hertz_fless_cgi_results} show the results obtained
by applying the CGI scheme for two values of spacing parameter $\kappa=2,3$ [see
Section~\ref{sec:cgi}]. The amplitude of
spurious oscillations are reduced for $\kappa=2$ ($\kappa<m_c$) and they are almost eliminated for
$\kappa=3$ ($\kappa=m_c$), enabling an accurate representation of the analytical contact
pressure distribution, only a small perturbation near edges of the contact zone is persisting. The results corroborate the
facts established by the patch and Eshelby tests \cite{akula_tying_paper}, concerning the minimum value
that $\kappa$ needs to take for minimizing the effect of mesh-locking
($\kappa\approx m_c$). Note that the change in vertical stress  in the MorteX
framework on the non-mortar side is simply a visualization issue due to inappropriate construction of contour plots in selectively integrated elements .
\begin{figure}[htb!]
   \centering
   \includegraphics[width=\textwidth]{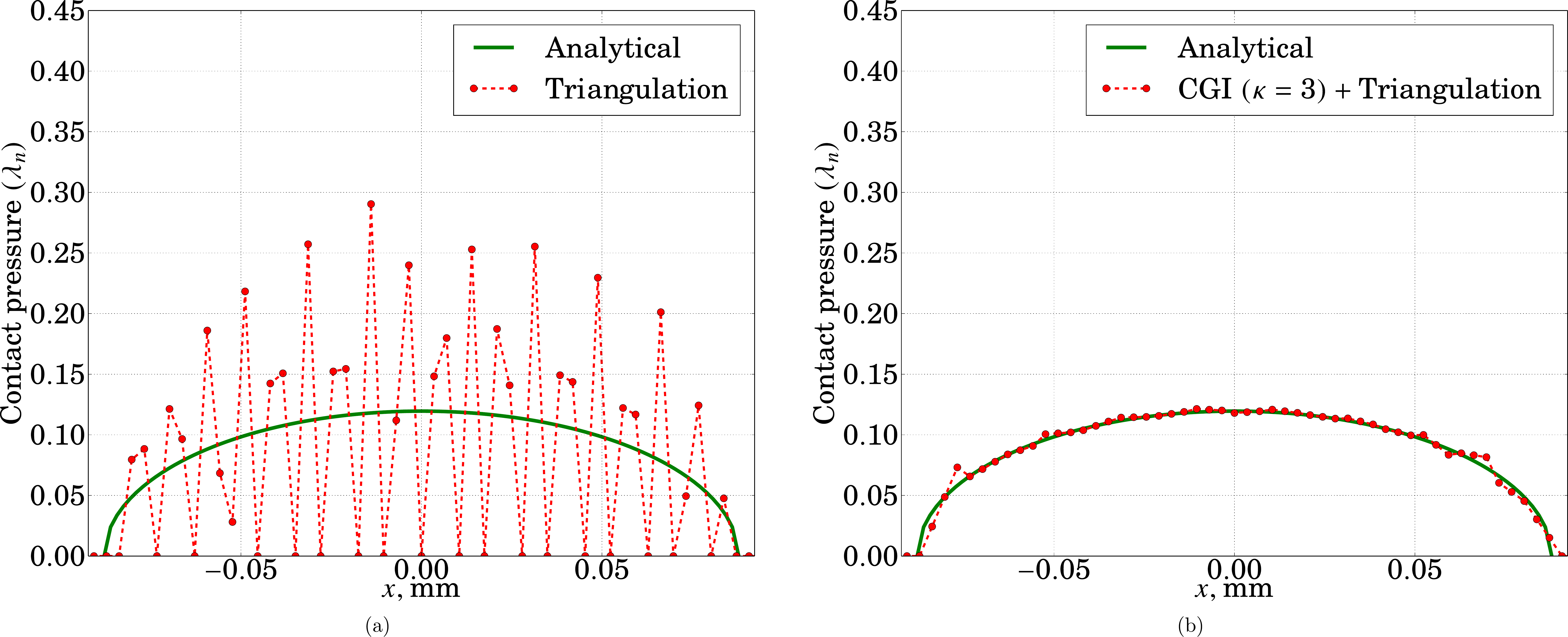}
   \caption{Contact tractions ($\lambda_n$) for the MorteX method (a) only
   triangulation of blending elements; (b) triangulation coupled with the CGI
   ($\kappa=3$).}
   \label{fig:cgi_triangulation_hertz}
\end{figure}
For the MorteX method, we additionally test the triangulation of blending
elements. As seen from Fig.~\ref{fig:cgi_triangulation_hertz}, the triangulation
of blending elements reduces the amplitude of oscillations (compared to
Fig.~\ref{fig:hertz_fless_sli_results}), which however remain significant. However,
when coupled with the CGI stabilization, they do not deteriorate the quality of
the solution.

\subsection{Frictional contact of cylinders}
A set-up similar to the frictionless case
[Fig.~\ref{fig:hertz_fricless_setup}(a)] is considered for the frictional contact.
The Coulomb's friction law is used
with the coefficient of friction $\mu = 0.2$.  The same linear elastic material is
assigned to the two cylinders with $E=200$ MPa and $\nu=0.3$. Both cylinders
are discretized ensuring equal meshes densities ($m_c\approx1$).
This test was also considered in~\cite{yang_two_2005,gitterle_finite_2010}. In the
first load sequence, a vertical displacement $u_y=0.182$ mm is applied on the
top surface of the upper cylinder in 100 load steps, which results in a total reaction force of
$P\approx 10.0$ N ($p_0\approx0.625$ N/mm). This is followed by a second sequence of loading where
a horizontal displacement $u_y=0.03$ mm is applied in 100 load steps. This results in a total
reaction force of $Q\approx 0.936$ N ($q_0\approx0.05851$ N/mm). 
The bottom surface of the lower cylinder is fixed throughout the simulation.
The contact pressure profile $p(x)$ and the contact semi-width $a$ are still
given by Eqs.~\eqref{eq:hertz_distribution}-\eqref{eq:hertz_con_width}.

According to the analytical solution~\cite{johnson_contact_1985}, the contact zone is divided
into a stick zone in the central area $|x|\leq c$ and two peripheral slip
regions $c<|x|\leq a$ , where the semi-width of the stick zone is given by:
\begin{equation}
    c=a\sqrt{1-\frac{q_0}{\mu p_0}}.
    \label{eq:hertz_fric_c}
\end{equation}
The tangential traction $p_\tau$ is given as:
\begin{equation}
 p_\tau(x) = \begin{cases}
           \quad\displaystyle\mu\frac{4Rp_0}{\pi a^2}(\sqrt{a^2-x^2}-\sqrt{c^2-x^2})&, \mbox{if}\,|x|\leq c;\\[12pt]
           \quad\displaystyle\mu\frac{4Rp_0}{\pi a^2}(\sqrt{a^2-x^2})&,\mbox{if}\,c<|x|\leq a;\\
           \quad 0&,\mbox{elsewhere.}
          \end{cases}
\end{equation}

\begin{figure}[htb!]
   \centering
   \includegraphics[width=\textwidth]{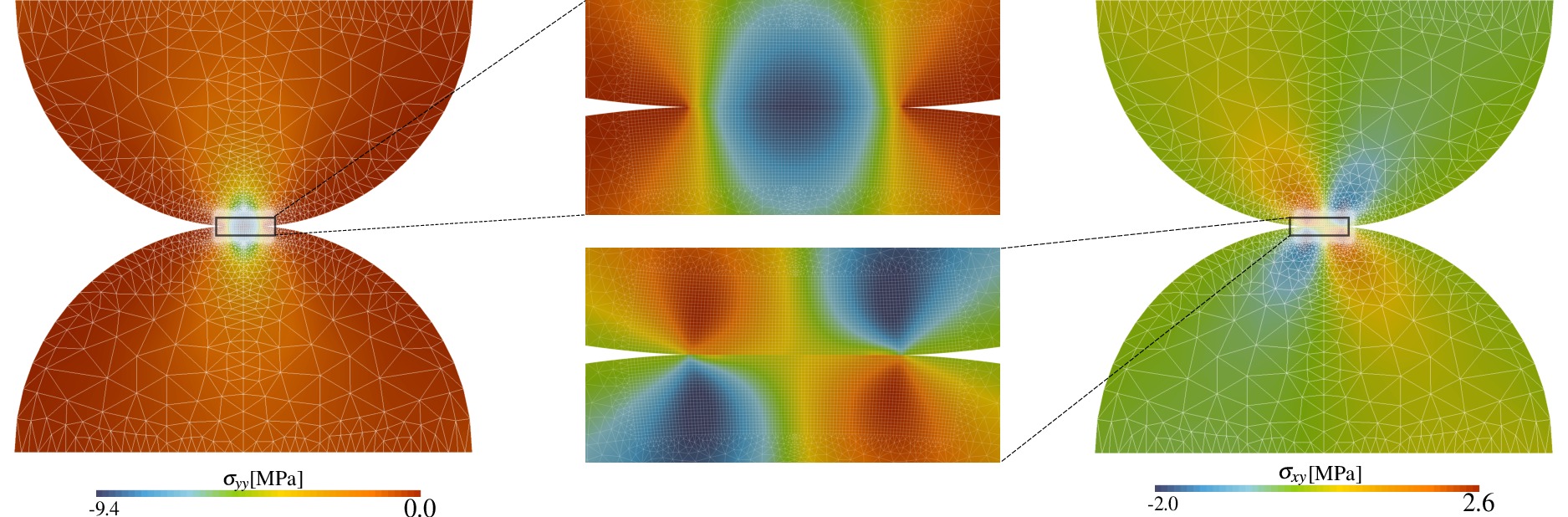}
   \caption{Contour stress plots $\sigma_{yy}$, $\sigma_{xy}$ for results
   obtained by Mortar method.}
   \label{fig:fric_hertz_results_mortar}
\end{figure}

\begin{figure}[htb!]
   \centering
   \includegraphics[width=\textwidth]{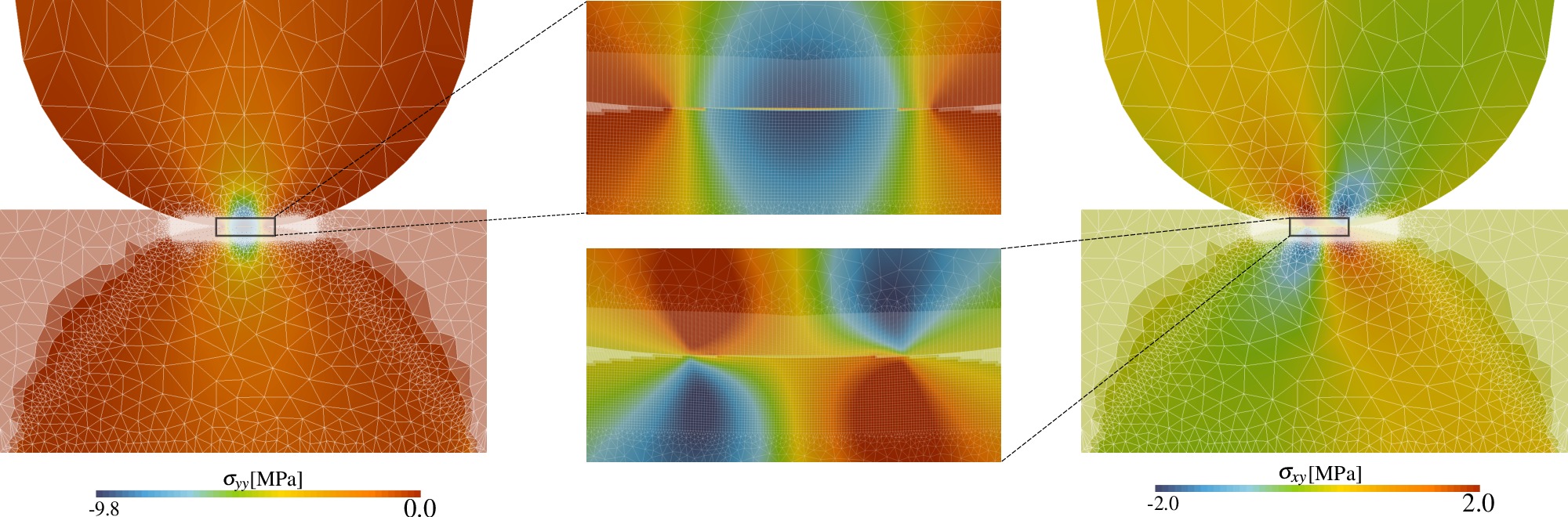}
   \caption{Contour stress plots $\sigma_{yy}$, $\sigma_{xy}$ for results
   obtained by MorteX method.}
   \label{fig:fric_hertz_results}
\end{figure}

\begin{figure}[htb!]
   \centering
   \includegraphics[width=\textwidth]{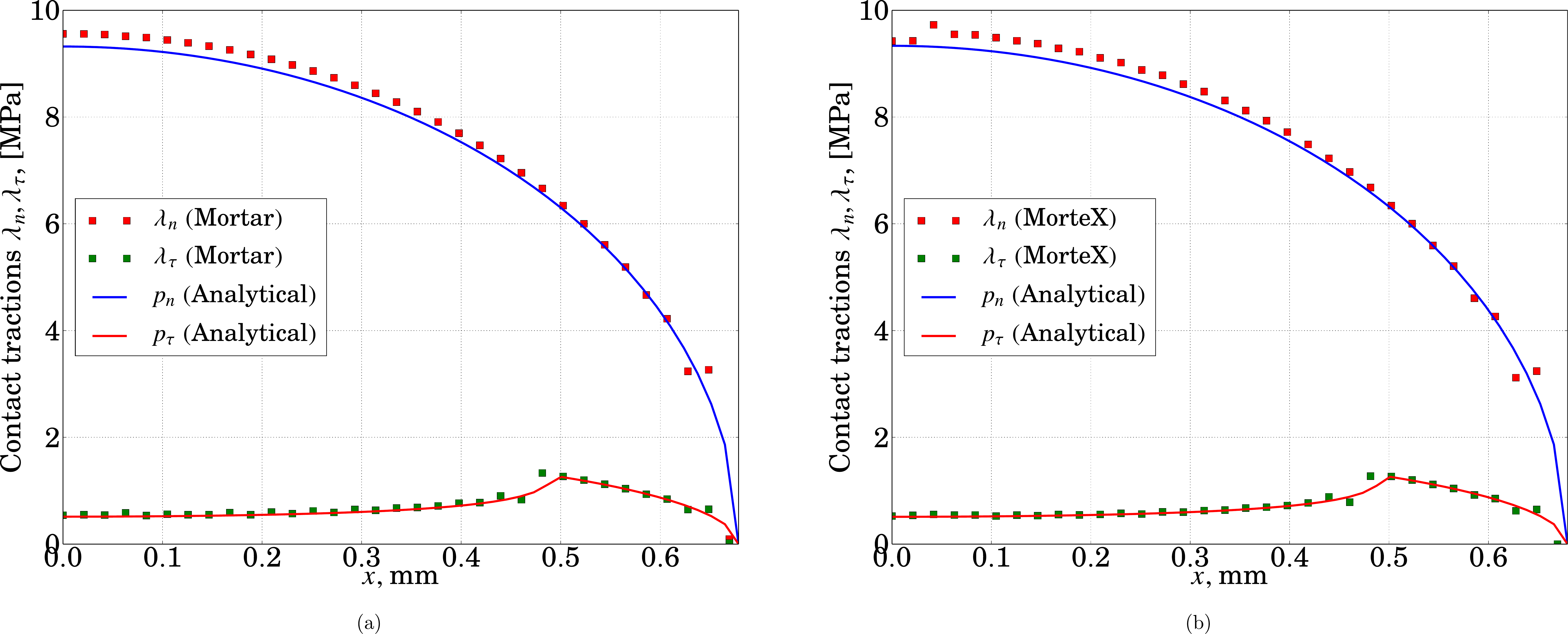}
   \caption{Normal and tangential contact tractions: (a) Mortar; (b) MorteX
   methods.}
   \label{graph:frictional_analytic}
\end{figure}

Fig.~\ref{fig:fric_hertz_results_mortar} and \ref{fig:fric_hertz_results}
show the contour stress plots for $\sigma_{yy}$ and $\sigma_{xy}$ at the end of
the second load sequence, obtained with
Mortar and MorteX methods, respectively. 
The results obtained by the Mortar and MorteX methods are very similar to each
other, and provide a good approximation of the analytical solution in terms
of contact tractions [see Fig.~\ref{graph:frictional_analytic}].

\subsection{Ironing a wavy surface}\label{subsec:fricless_ironing}
\begin{figure}[htb!]
   \centering
   \includegraphics[width=.85\textwidth]{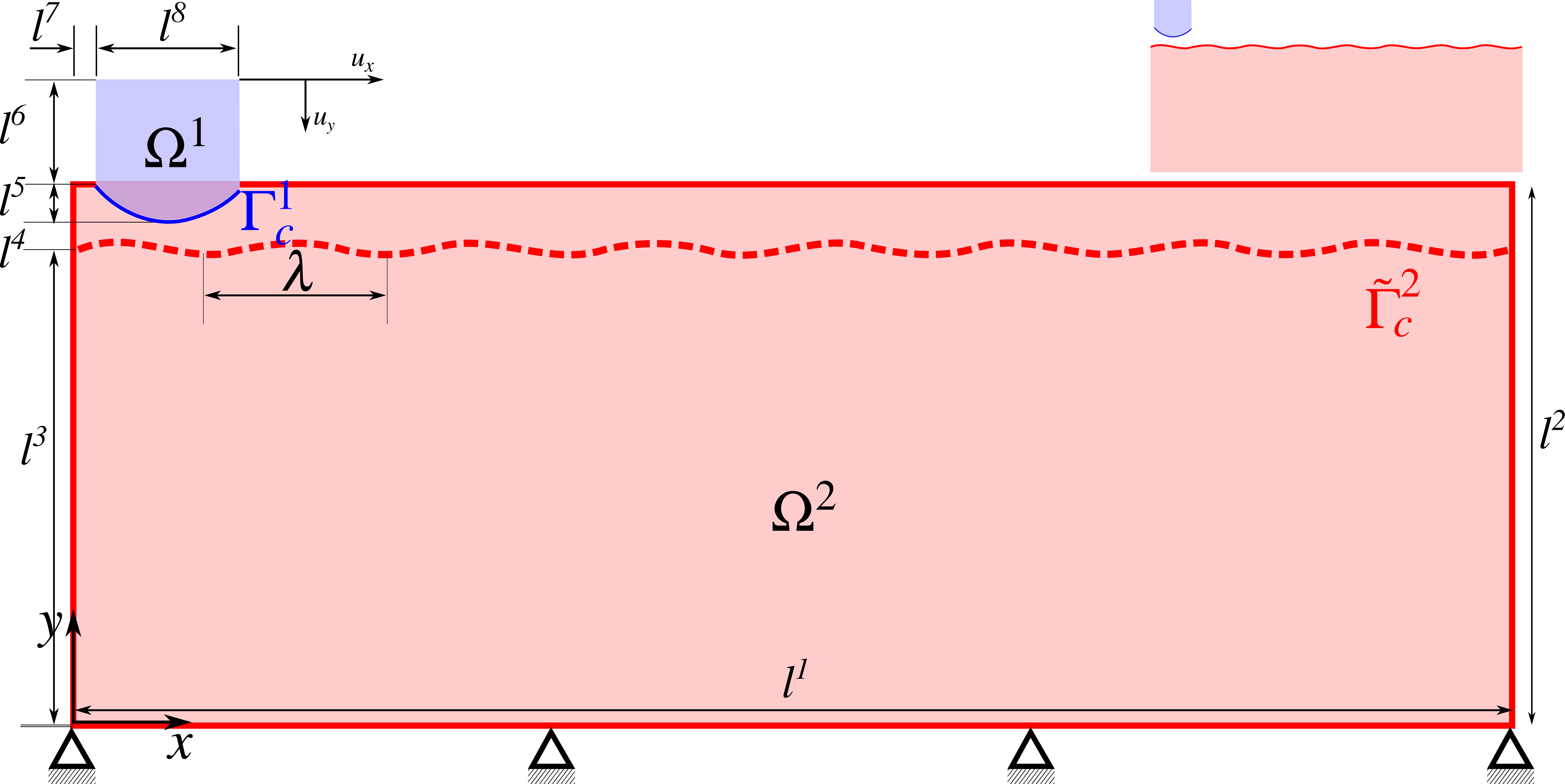}
   \caption{MorteX frictionless ironing of wavy surface set-up, equivalent
   Mortar set-up (inset).}
   \label{fig:wavy_iron_setup}
\end{figure}
In this example we consider a frictionless sliding contact between an elastic
slider and an elastic wavy substrate [see
Fig.~\ref{fig:wavy_iron_setup}].  Within the MorteX framework, a
discretized surface $\Gamma_{c}^1$ comes in contact and slides along a virtual
surface $\tilde\Gamma_c^2$ embedded inside a rectangular block. The slider and substrate
are meshed with a comparable mesh density ($m_c\approx 1$).
The geometric dimensions used are: $l^1=12,\,l^2=4.5,\,l^3=4,\,l^4=0.2,\,l^5=0.3,\,l^6=0.9,\,l^7=0.2,\,l^8=1.2$
(all the length dimension are in mm). The wavy surface is described by
$y(x)=l^3+\Delta\sin(2\pi x/\lambda)$, with $\lambda=1.5$ mm and
$\Delta=0.05$ mm.  Both solids
are made of the same material: Young's modulus $E=100$ GPa and Poisson's ratio
$\nu=0.3$.  A vertical displacement of $u_y=-0.75$ mm is applied on the top of
the slider within first 20 load steps ($t\,\in[0,1]$) while the horizontal displacement of the
same boundary is kept zero. During the following sequence ($t\,\in[1,2]$), the vertical
displacement is maintained and the horizontal displacement of $u_x=10$ mm is
applied in 100 load steps. The bottom surface of the rectangular block is
fixed throughout the simulation.

\begin{figure}[htb!]
   \centering
   \includegraphics[width=\textwidth]{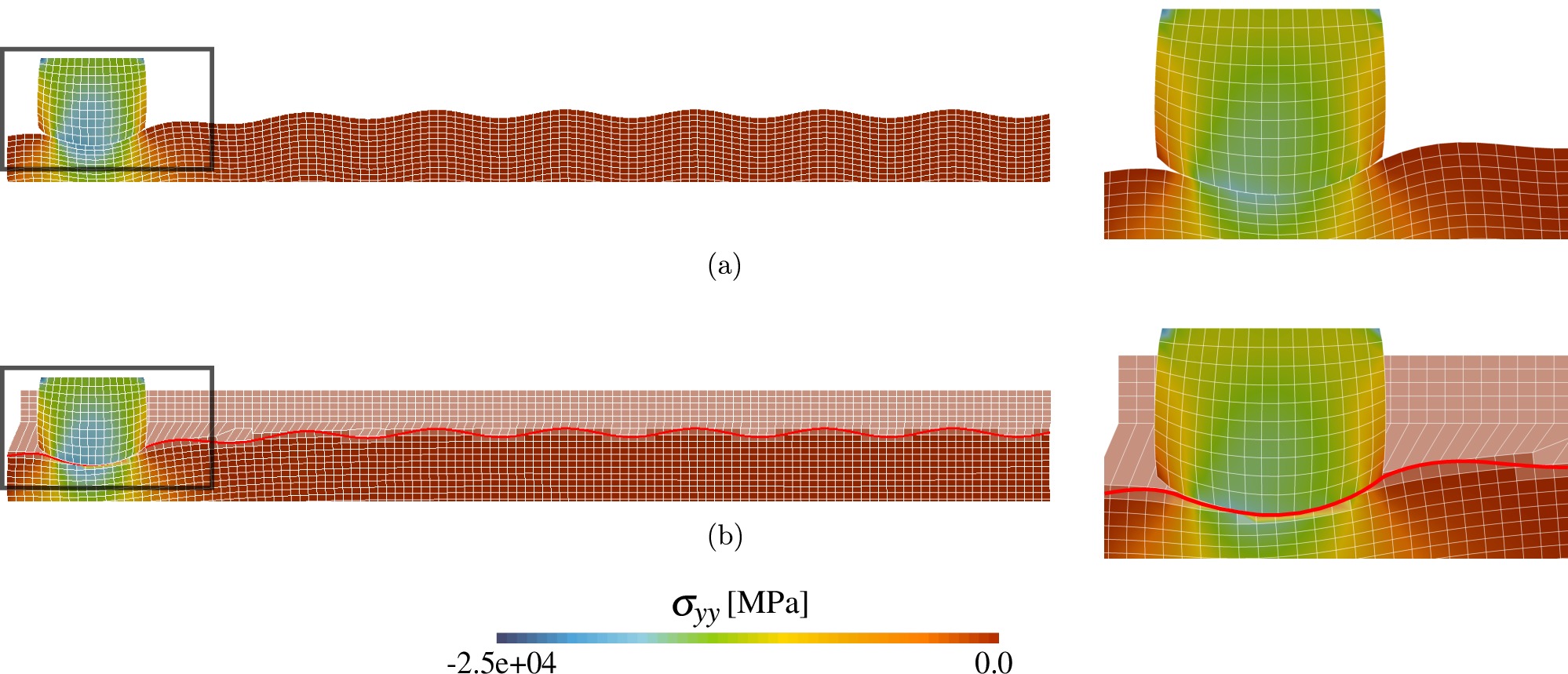}
   \caption{$\sigma_{yy}$ contour plots at $t=1$ s (a) Mortar; (b)
   MorteX methods.}
   \label{fig:wavy_stresses_mortar}
\end{figure}

\begin{figure}[htb!]
   \centering
   \includegraphics[width=\textwidth]{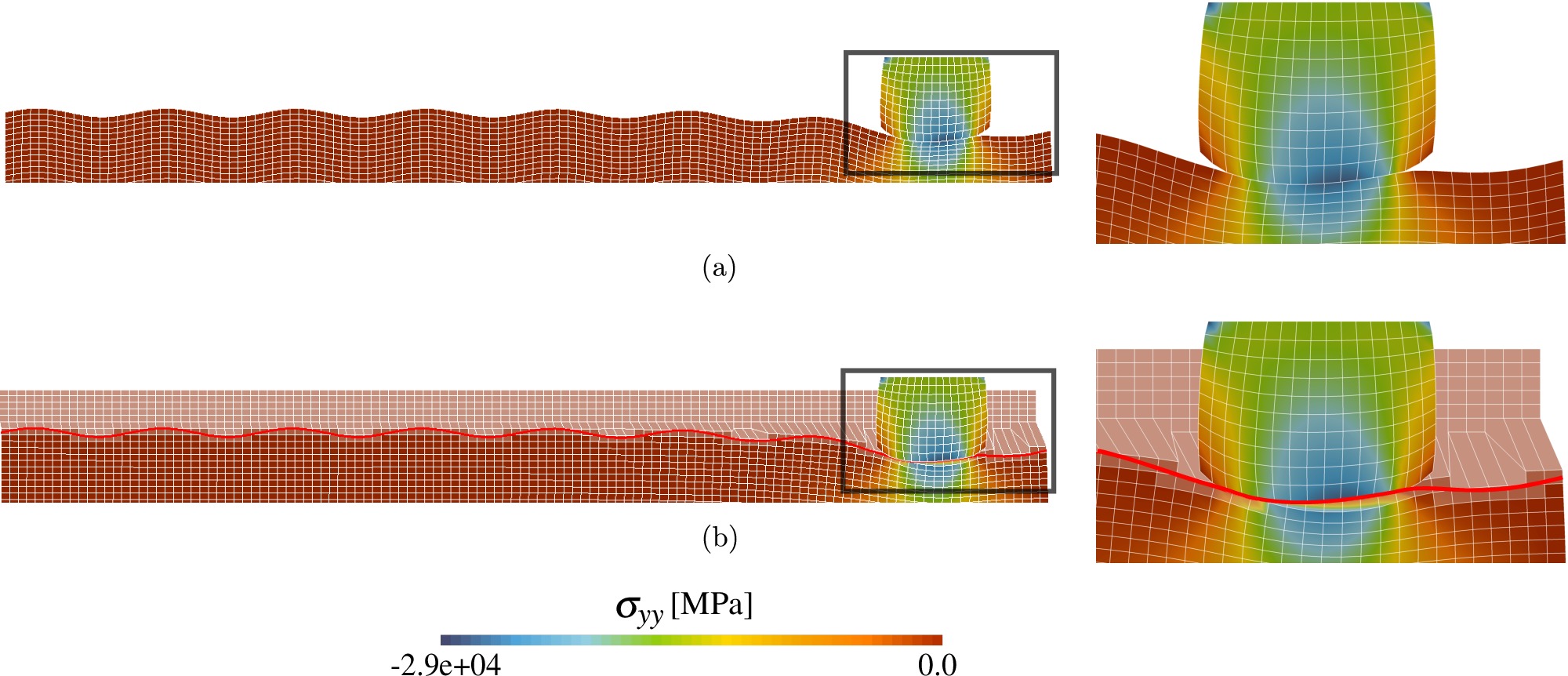}
   \caption{$\sigma_{yy}$ contour plots at $t=2$ s (a) Mortar; (b)
   MorteX methods.}
   \label{fig:wavy_stresses_morteX}
\end{figure}

\begin{figure}[htb!]
   \centering
   \includegraphics[width=\textwidth]{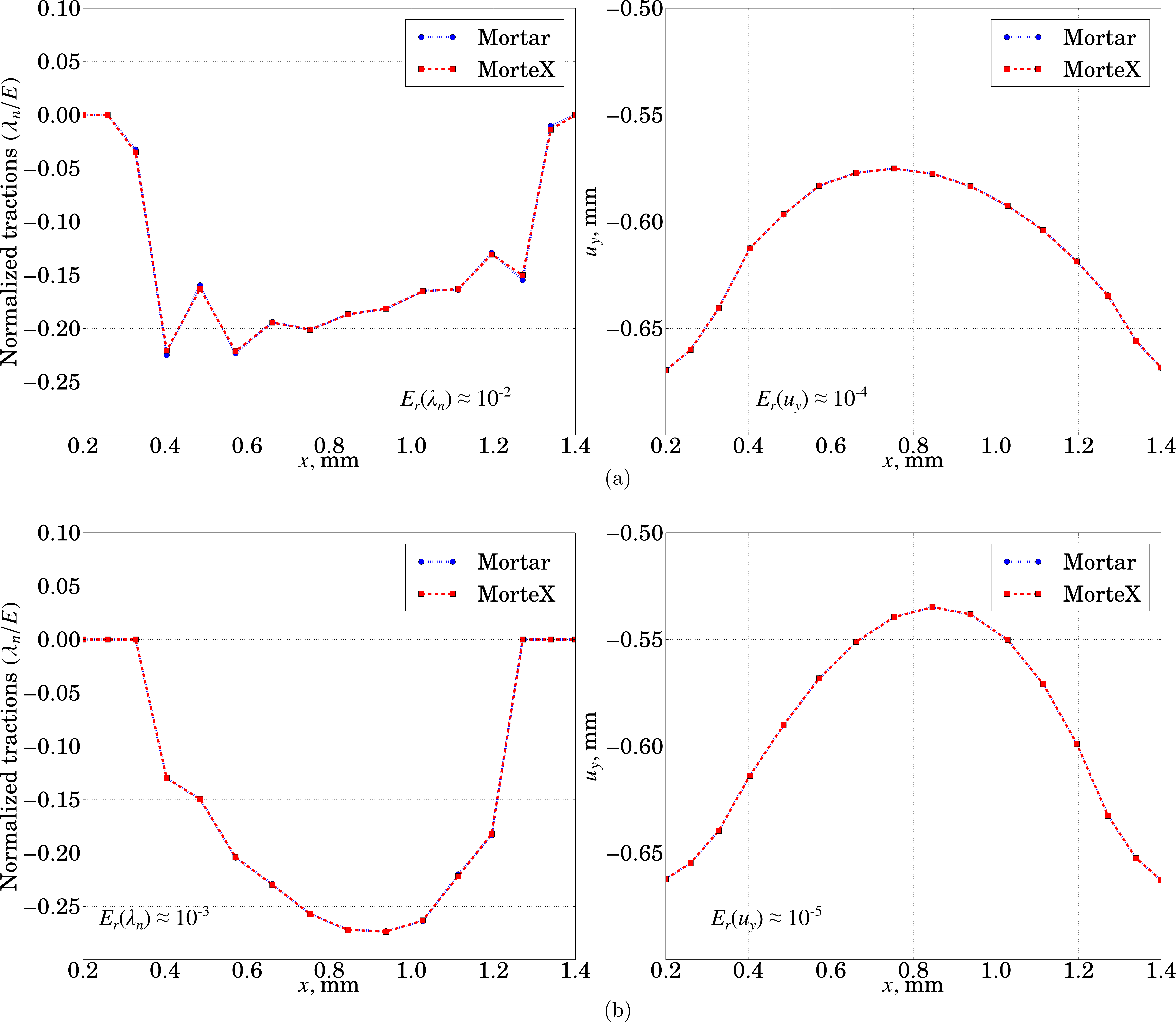}
   \caption{Comparison of normalized contact tractions ($\lambda_n/E$) and
   displacements ($u_y$) along slider surface $\Gamma_c^1$: (a) $t=1$ s; (b) $t=2$ s; associated norms of difference between displacements and pressures obtained by the MorteX and Mortar methods are also shown.}
   \label{fig:wavy_lam_U_plots}
\end{figure}

For the sake of comparison, the same problem is also solved within the classical mortar
contact formulation, in which the contact occurs between two surfaces explicitly
represented by body-fitted meshes [see Fig.~\ref{fig:wavy_iron_setup}, inset].  
The contour plots of stresses $\sigma_{yy}$, at $t=1.0,2.0$ seconds, for both the Mortar and MorteX methods are
shown in Fig.~\ref{fig:wavy_stresses_mortar}-\ref{fig:wavy_stresses_morteX}.
These fields are very smooth and indistinguishable by naked eye. 
The normalized contact tractions ($\lambda_n/E$) and the displacements in
$y$-direction ($u_y$) along the real surface $\Gamma_{c}^1$ are compared between
the two methods.  In order to quantify the difference between two solutions we use $L^2$
norm of displacement and contact traction difference as below:\\
\begin{tabularx}{\textwidth}{XX}
\begin{equation}
E_r (u_y) = \frac{|| u_y^{\mbox{\tiny Mortar}}- u_y^{\mbox{\tiny
MorteX}}||_{L^2(\Gamma_c^1)}}{|| u_y^{\mbox{\tiny
Mortar}}||_{L^2(\Gamma_c^1)}},
\label{eq:lam_norm}
\end{equation}&
\begin{equation}
E_r(\lambda_n) = \frac{||\lambda_n^{\mbox{\tiny Mortar}}-\lambda_n^{\mbox{\tiny
MorteX}}||_{L^2(\Gamma_c^1)}}{||\lambda_n^{\mbox{\tiny
Mortar}}||_{L^2(\Gamma_c^1)}}.
\label{eq:lam_norm}
\end{equation}
\end{tabularx}
where the norm is computed as $\left\|f(x)-g(x)\right\|_{L^2(\Gamma_g^1)} = \sqrt{\frac 1 N\sum_{i=1}^N
 \left[f(x_i)-g(x_i)\right]^2}$, where $x_i \in [0,L]$ are the $x$-coordinates of
 mortar nodes $\vec x_i \in\Gamma_g^1$ in the reference configuration.
Negligibly small difference between displacements and tractions obtained by the two methods [see Fig.~\ref{fig:wavy_lam_U_plots}] 
reflects a comparable accuracy the MorteX framework compared to the classical Mortar method.


\subsection{Frictional shallow ironing}
\begin{figure}[htb!]
   \centering
   \includegraphics[width=.85\textwidth]{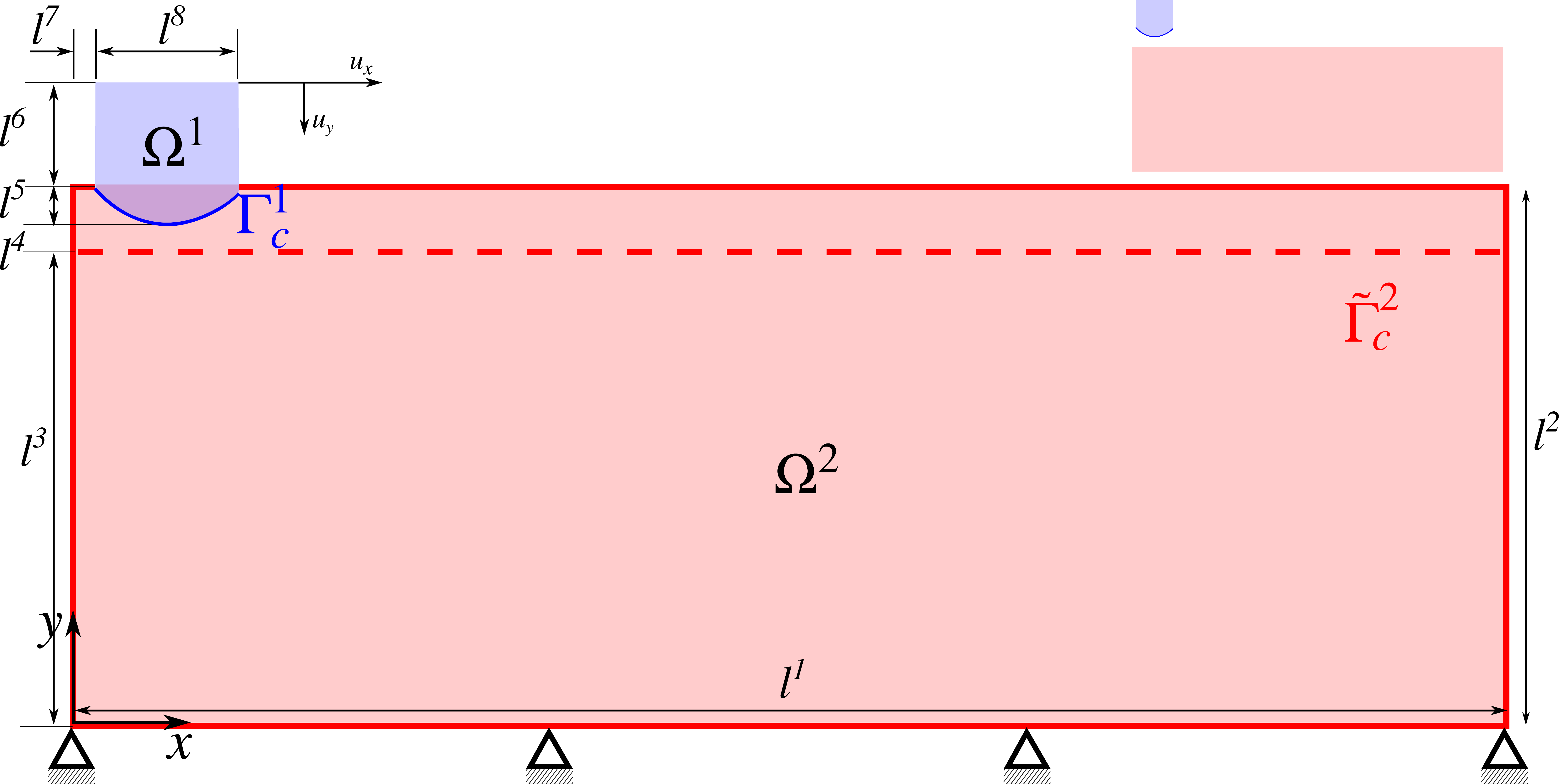}
   \caption{MorteX frictional ironing set-up, equivalent Mortar set-up (inset).}
   \label{fig:fric_iron_setup}
\end{figure}
Here we consider the same geometrical set-up as in the previous example
(Section~\ref{subsec:fricless_ironing}), but here the
virtual surface $\tilde\Gamma_c^2$ embedded into the host domain $\Omega^2$
is flat [see Fig.~\ref{fig:fric_iron_setup}].  The slider and substrate properties are respectively
$E_1=68.96$ MPa, $\nu_1=0.32$ and $E_2=6.896$ MPa, $\nu_1=0.32$ ($E_1/E_2=10$). 
In addition, the slider has a finer mesh than the
substrate: so that the mesh contrast is $m_c\approx 3$. A
coefficient of friction $\mu=0.3$ is used.  The contrast in material and mesh
density is introduced purposefully to better illustrate the manifestation of
the mesh locking phenomenon.  A vertical displacement of $u_y=-0.75$ mm is
applied on the top of the slider within the first 50 load steps ($t\,\in[0,1]$),
while the horizontal displacement of the same boundary is kept zero. During the
    following sequence ($t\,\in[1,2]$), the vertical displacement is maintained
    and the horizontal displacement of $u_x=10$ mm is applied in 500 load
    steps. The bottom surface of the rectangular block is fixed in all directions throughout the simulation.

\begin{figure}[htb!]
   \centering
   \includegraphics[width=.85\textwidth]{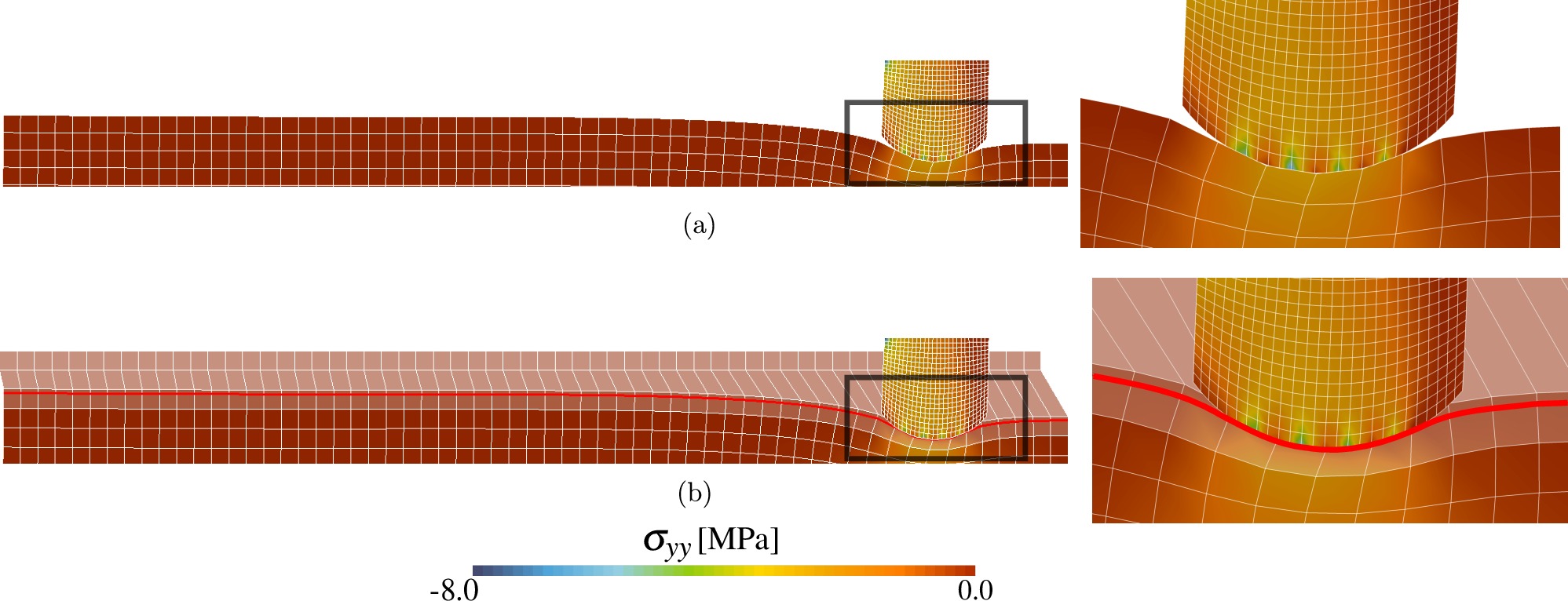}
   \caption{Standard Lagrange multiplier space ($\kappa=1$), $\sigma_{yy}$
   contour plots at $t=2$ seconds (a) Mortar; (b) MorteX methods.}
   \label{fig:fric_iron_contour_spacing_1}
\end{figure}

\begin{figure}[htb!]
   \centering
   \includegraphics[width=.85\textwidth]{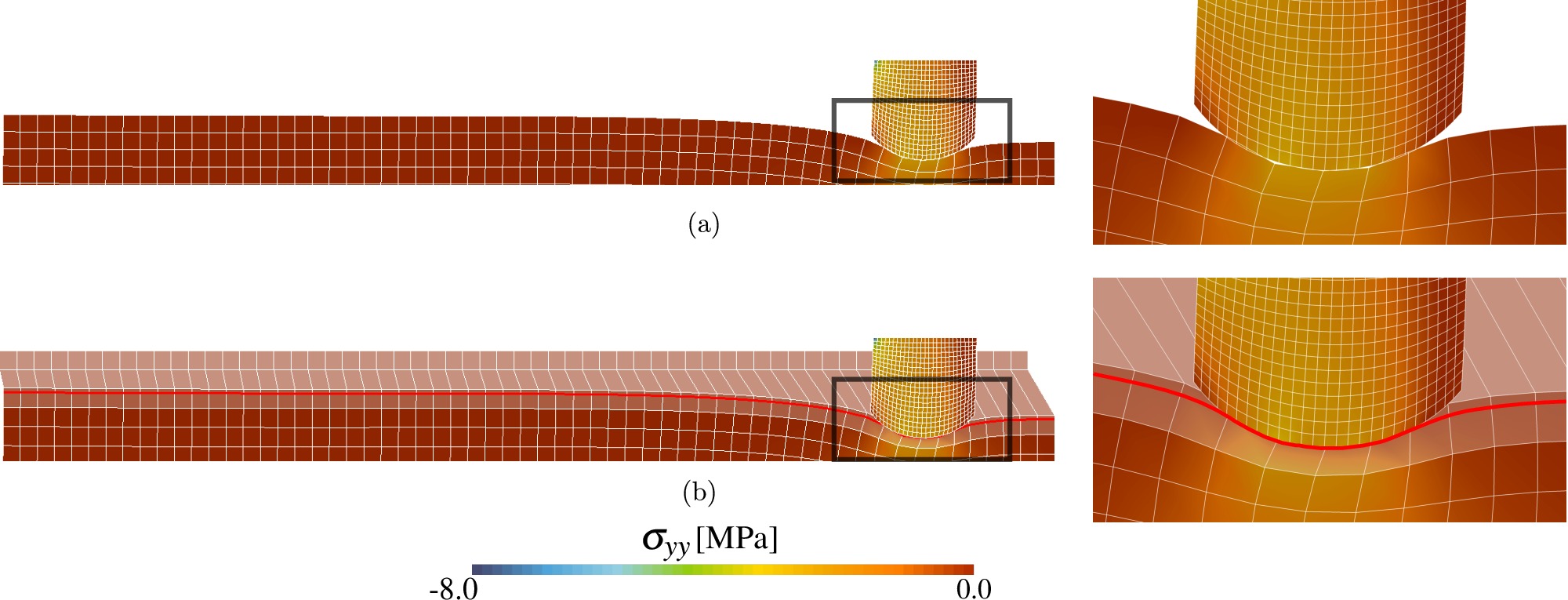}
   \caption{Coarse grained Lagrange multiplier space ($\kappa=3$), $\sigma_{yy}$
   contour plots at $t=2$ seconds (a) Mortar; (b) MorteX methods.}
   \label{fig:fric_iron_contour_spacing_2}
\end{figure}

\begin{figure}[htb!]
   \centering
   \includegraphics[width=\textwidth]{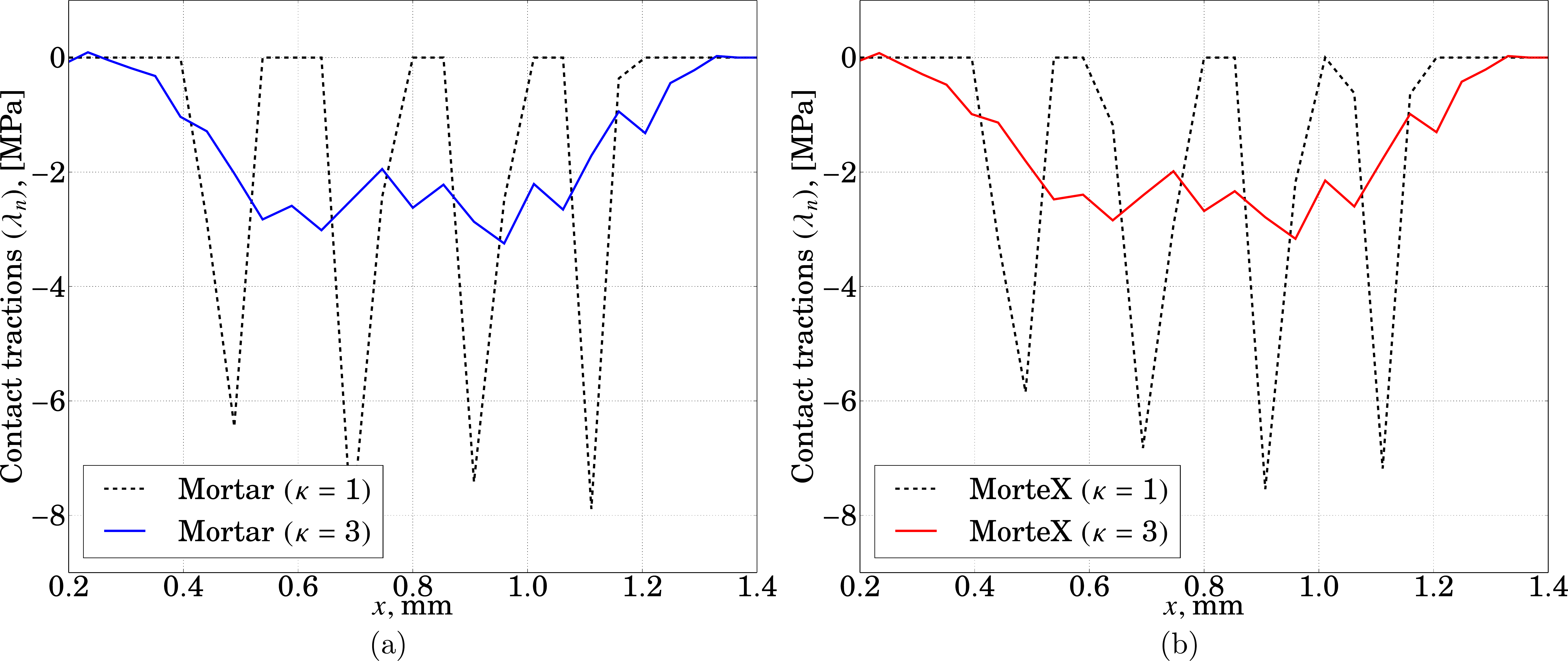}
   \caption{Comparison of $\lambda_n$ at nodes along $\Gamma_c^1$ for $\kappa=1$
   and $\kappa=3$: (a) Mortar; (b) MorteX methods at time $t=2$ s.}
   \label{fig:fric_iron_spacing_plots}
\end{figure}

\begin{figure}[htb!]
   \centering
   \includegraphics[width=\textwidth]{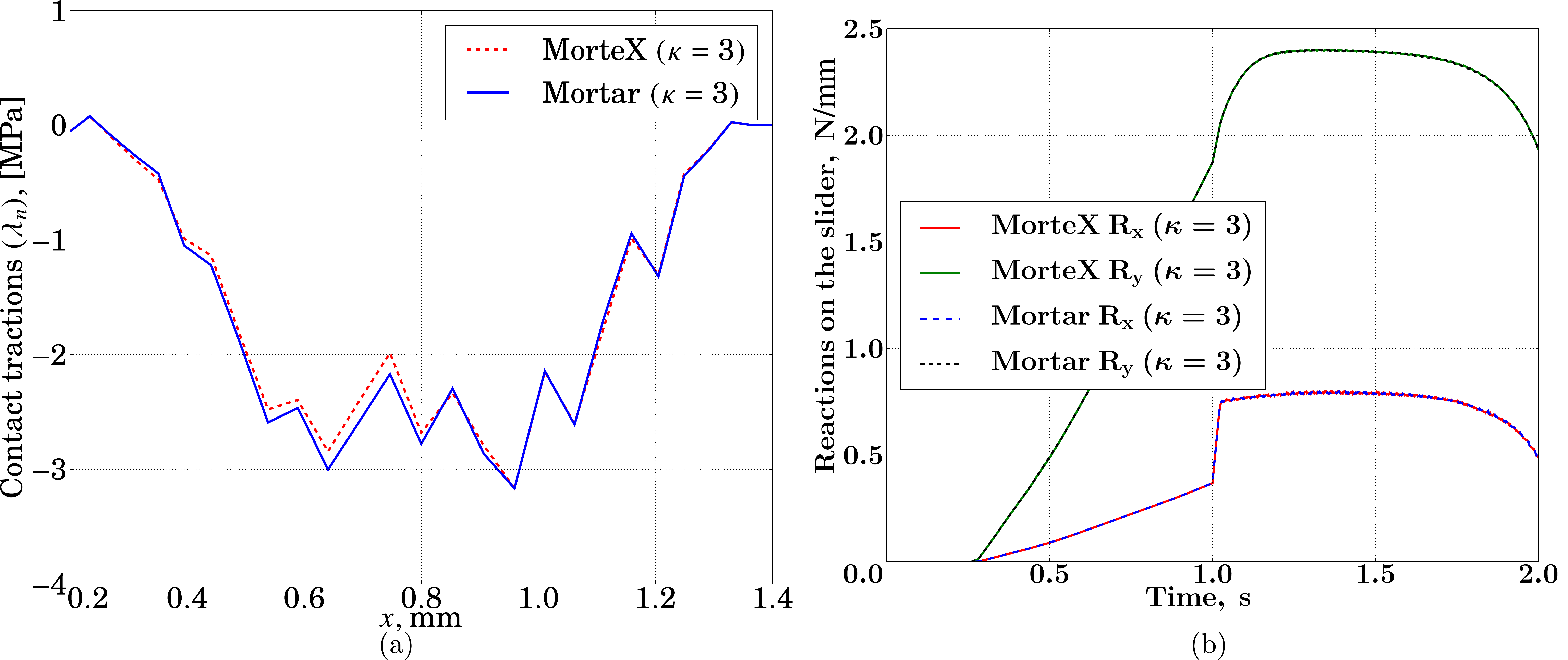}
   \caption{Comparison of  Mortar and MorteX methods for CGI with
   $\kappa=3$: (a) contact tractions $\lambda_n$ along $\Gamma_c^1$; (b)
   reaction forces in $x$ and $y$ along the top surface of the slider.}
   \label{fig:fric_iron_kappa3_results}
\end{figure}

Fig.~\ref{fig:fric_iron_contour_spacing_1}, shows the oscillation in the stress
field $\sigma_{yy}$ for the standard Lagrange multiplier interpolations, both in the
context of Mortar and MorteX. These results suffer from spurious oscillations.  Similar to
the Hertzian contact problem setting in Section~\ref{subsec:hertz_sol}, we
use the coarse grained Lagrange multiplier interpolations to both the Mortar and
MorteX formulations. It results in a reduced amplitude of oscillations, as
seen in
Fig.~\ref{fig:fric_iron_contour_spacing_2} and
\ref{fig:fric_iron_spacing_plots} where the contact tractions are presented for
different coarse grain spacing parameter $\kappa$.
A good agreement between the Lagrange multiplier
field $\lambda_n$ and hence the reaction forces ($R_x, R_y$) on the slider for the
Mortar and MorteX methods can bee seen in Fig.~\ref{fig:fric_iron_kappa3_results}.
In this example, the applicability of CGI for Lagrange multipliers is once again
proved efficient for both the MorteX and for the classical
Mortar frameworks.

\section{Concluding remarks\label{sec:conclusion}}

The MorteX framework combining the selective integration from the X-FEM and accurate handling of contact constraints between a boundary fitted and embedded surfaces is developed for the two-dimensional case.
The use of the monolithic augmented Lagrangian method renders the constrained optimization problem (resulting from frictional contact) to a fully unconstrained one and ensures a robust convergence because of the resuting smoothed functional which does not require the use of active set strategy common to the method of Lagrange multipliers. Moreover, it presents a comprehensive framework to solve frictional problems without violation of stick constraints inherent to the penalty method. However, for high mesh-density and material contrast, the resulting mixed formulation is shown to be prone to over-constraining of the interface or mesh locking which manifests itself in the form of spurious oscillations. 
As demonstrated by the classical patch test, these oscillations can be removed if triangular elements are used in the mesh hosting an embedded surface (at least the blending element should be made triangular).
However, in real applications, such simple triangulation cannot ensure oscillation-free interface traction fields. To deal with this mesh locking, we adopt here the technique of coarse-grained interpolation of Lagrange multipliers developed in~\cite{akula_tying_paper}. It consists in assigning Lagrange multipliers only to few mortar-side nodes (master nodes); the choice of the subsampling distance is governed by the relative mesh densities, i.e. the local or global ratio of active mortar nodes per number of active blending elements in contact. The value of Lagrange multipliers of the remaining mortar (slave) nodes is ensured by shape functions spanning few elements. Alternatively, Lagrange multipliers can be assigned to all mortar nodes, but those of the slave nodes can be constrained using multi-point constraints and associated interpolation between master nodes. Such coarse graining cannot ensure a perfect performance of the contact patch test but it enables to reduce the amplitude of spurious oscillations by more than two orders of magnitude for large material contrast (ratio of Young's moduli equal to 1000) and high mesh density contrast (ratio of the number of mortar nodes per number of blending elements equal to 10). For more realistic contact problems such as frictional contact between two cylinders or shallow ironing, the coarse graining ensures accurate results free of spurious oscillations. Moreover, this coarse-graining is shown to be helpful for the classical mortar method applied to contact problems between boundary fitted surfaces. As demonstrated here, if the material and mesh-density contrast are high between contacting surfaces, spurious oscillations also manifest themselves and can be efficiently removed by the coarse-graining technique. To conclude, we would like to highlight that the good performance of the discretization scheme in the classical contact patch test, where a uniform pressure is transmitted in the contact interface,  does not ensure a good performance in a general case. Therefore, it is suggested to extend the contact patch test by applying not only a uniform pressure but also a bending moment~\cite{sanders_nitsche_2012,akula_tying_paper}.

The developed framework should be helpful for wear simulations, where the worn surface can be assumed to be embedded and can propagate inside the bulk mesh due to material loss. The current work is focused on elaborating a methodology to simulate wear within the MorteX framework with an energy-based local wear law. In perspective, this framework is planned to extend to the three-dimensional case. This extension should also involve the possibility to coarse grain Lagrange multipliers and use an appropriate interpolation spanning several mortar faces.

\section*{Acknowledgment}
The authors acknowledge financial support of the ANRT (grant CIFRE no
2015/0799). We are also grateful to Nikolay Osipov and St\'ephane Quilici from the
Z-set team for their help with the implementation of the method.



\end{document}